\documentclass[preprint,12pt]{elsarticle}
\usepackage[table]{xcolor}
\usepackage{lscape}
\usepackage{amssymb}
\usepackage{amsmath}
\usepackage{placeins}
\usepackage{footnote}
\usepackage{tabulary}
\usepackage{graphicx}
\usepackage{graphics}
\usepackage{caption}
\usepackage{subcaption}
\usepackage{booktabs}% http://ctan.org/pkg/booktabs

\usepackage{enumitem}
\journal{arXiv}
\begin{document}

\begin{frontmatter}

\title{Assessing the predictive ability of the UPDRS
for falls classification in early stage Parkinson's disease}

\author{Sarini Abdullah$^1$, Nicole White$^2$, James McGree$^3$, Kerrie Mengersen$^3$, Graham Kerr$^2$}
\address{$^1$ Department of Mathematics, University of Indonesia\\ $^2$Institute of Health and Biomedical Innovation (IHBI), Australia\\$^3$ARC Centre of Excellence for Mathematical and Statistical Frontiers, Queensland University of Technology (QUT), Australia}

\begin{abstract}

\textbf{Background:} Falling is common for people with Parkinson's disease (PD), with negative consequences in terms of quality of life. Therefore, the identification of risk factors associated with falls is an important research question. In this study, various ways of utilizing the Unified Parkinson's Disease Rating Scale (UPDRS) were assessed for the identification of risk factors and for the prediction of falls.

\textbf{Methods:} Three statistical methods for classification were considered: decision trees, random forests, and logistic regression. For variable selection, the stepwise procedure and Bayesian model averaging based on log-marginal likelihood were implemented for logistic regression, and Gini index criterion was used for decision trees and random forests. UPDRS measurements on 51 participants with early stage PD, who completed monthly falls diaries for 12 months of follow-up were analyzed. 

%\justify
\textbf{Results:} All classification methods applied produced similar results in regards to classification accuracy and the selected important variables. The highest classification rates were obtained from model with individual items of the UPDRS with 80\% accuracy (85\% sensitivity and 77\% specificity), higher than in any previous study. A comparison of the independent performance of the four parts of the UPDRS revealed the comparably high classification rates for Parts II and III of the UPDRS. Similar patterns with slightly different classification rates were observed for the 6- and 12-month of follow-up times. Consistent predictors for falls selected by all classification methods at two follow-up times are: thought disorder for UPDRS I, dressing and falling for UPDRS II, hand pronate/supinate for UPDRS III, and sleep disturbance and symptomatic orthostasis for UPDRS IV. While for the aggregate measures, subtotal 2 (sum of UPDRS II items) and bradykinesia showed high association with fall/non-fall.

\textbf{Conclusions:}  Fall/non-fall occurrences were more associated with individual items of the UPDRS than with the aggregate measures. UPDRS parts II and III produced comparably high classification rates for fall/non-fall prediction. Similar results were obtained for modelling data at 6-month and 12-month follow-up times.
\end{abstract}


\end{frontmatter}

\section{Introduction}
\label{S:1}

Falls are a significant and common problem for persons diagnosed with Parkinson’s disease (PD)\cite{Balash2005, Roller1989,  Wood2002}, and are prominent even in the disease's early stages \cite{Bloem2001, Kerr2010}.  Several prospective studies have shown that falls incidence is relatively high among people with Parkinson's (PWP), with estimates ranging from 46-72\%, over three, six and twelve month periods \cite{Bloem2001,Gray2000, Kerr2010, Pickering2007, Wood2002}. PD fallers are also more likely to fall in the future \cite{Pickering2007, Wood2002}. The negative consequences of falls for PWP quality of life \cite{Bloem2001,Karlsen2000, Thomas2008} and associated health care costs \cite{FranAois2017} combined with its high prevalence has motivated the investigation of risk factors associated with their occurrence.

Prospective falls methods are the gold standard, in contrast to retrospective falls, for falls prediction and falls risk factors identification. The problem with looking at retrospective falls is that elderly may forget they have fallen. Despite this gold standard, there are relatively few prospective falls studies. Among the few are as in \cite{Wood2002, Kerr2010, Gazibara2015}. \cite{Kerr2010} pointed out the inconsistency in clinically useful falls risk factors in 7 prospective studies. In search for the effective way to predict falls, \cite{Kerr2010} put more attention to the functional tests and disease-specific clinical assessments and developed a multivariate predictive model. It is inferred that a combination of both disease-specific and balance-and mobility-related measures can accurately predict falls in PWP. 

The Unified Parkinson’s Disease Rating Scale (UPDRS) and its recent revision, Movement Disorder Society-UPDRS (MDS-UPDRS) \cite{Goets2008}, is considered to be the gold standard instrument for the clinical assessment of PD \cite{Song2009}, due to its high reliability \cite{MartinezMartin1994, MDS2003} and thorough measurement of multiple factors including body structure and function, activity and participation \cite{Song2009}.  For these reasons, the UDPRS has been utilised by a number of studies for falls prediction, as an overall measure of disease severity \cite{Drapier2005, Fahn2004, Kerr2010, Moro2010,  Pickering2007, Wood2002}.  These studies have consistently shown positive associations between falling and higher UPDRS scores, namely Part II and III subtotals and the overall (Part I-IV) sum. In addition, sums of different combination of UPDRS individual items are often obtained to give a single measure for a symptom, such as: tremor, rigidity, bradykinesia, and Postural Instability and Gait (PIGD). These composite measures also showed a reasonably positive association with falls. The use of these aggregate measures, however, ignores the contribution of individual UPDRS items to the prediction of falls risk, many of which are functionally relevant and potentially managed.  

Logistic regression is a popular statistical tool for classification as a function of observed predictors.  It has been extensively used for falls classification in PD \cite{Allcock2009, Kerr2010, Pickering2007, Sun2009, Wood2002}.  The popularity of logistic regression is due to its ease of interpretation via odds ratios and the estimation of individual patient probabilities into faller and non-faller groups, for the purpose of deriving an optimal classification rule.  However, the determination of the best subset of predictors for inclusion in logistic regression is challenging and, for this reason, analysis is often restricted to first-order (or linear) effects. Higher order effects, for example the interaction between predictors is ignored. Moreover, one guideline suggests that there should be at least 10 participants for each predictor \cite{Agresti2007} which is often difficult to fulfill in many PD data applications. 

Bayesian model averaging (BMA) offers an appealing solution to optimal model selection by combining predictions from multiple models, with different subsets of predictors \cite{Hoeting1999}.  This technique applies concepts from Bayesian inference by weighting different models according to their posterior model probability and producing a consensus prediction for the outcome of interest.  The appeal of BMA is driven by accounting for model uncertainty which here relates to whether covariates, interactions and high order terms should appear in the model. Although several health-related studies have implemented this approach \cite{Viallefont2001, Wang2004, Fang2016}, 
to the best of our knowledge, this paper is the first to implement BMA for Parkinson's related study.

Decision trees (DTs) \cite{Quinlan1986} and random forests (RFs) \cite {Breiman2001} are examples of nonparametric tree-based classification methods that naturally incorporate variable selection. The natural variable selection is owing to the tree construction mechanism employed in easy method. A tree is grown by first selecting the most discriminating variable (called splitting variable) to partition the data into the target classes (child nodes). Then, the process is repeated in each of the child nodes, until a certain stopping criterion is reached. Thus, only important variables are used in decision making about the predicted classes.   

Both methods employ recursive partitioning to automatically determine predictors that best discriminate between classes of the response variable, resulting in tree-like structures. The very nature of these tree-like structures accommodate complex interactions without suffering from the curse of dimensionality \cite{Friedman1997a}. They have also been popularized due to their ease of interpretation. Using these methods, the aim of this paper was to evaluate the utility of the UPDRS for falls classification in people with early stage PD, with a view to identifying key predictors that contribute to this classification.  This will provide useful information to clinicians as a more focused attention to the identified factors could provide a better guide in understanding the patients condition. Moreover, a quick and straightforward decision on the likely of falls in patients could be inferred using the ''if then rules'' in decision trees. 

The remainder of this paper is organized as follows.  A description of the data and methodology are provided in Section~\ref{Sec:method}.  Key results are presented in Section~\ref{Sec:result}, including the comparison of individual UPDRS items versus composite measures, the relative importance of different UDPRS subsections, and the identification of key predictors.  A discussion of results and limitations are presented in Section~\ref{Sec:discussion} and a summary of overall findings is given in Section~\ref{Sec:summary}.

\section{Data and Methods}
\label{Sec:method}
\subsection{Participants}

Fifty one participants diagnosed with idiopathic PD were recruited for this study, as part of a larger research project conducted by the Institute of Health and Biomedical Innovation in Brisbane, Australia \cite{Kerr2010}.  All participants were classified as early stage PD, determined by a Hoehn and Yahr (HY) score of 3 or less.

Each participant completed a monthly falls diary over a consecutive period.  Based on this information, a participant was classified as a faller if they had experienced at least one fall within a defined follow-up period.  In this study, follow-up times were defined at six and twelve months.  Successful completion of the diary was monitored by phone calls and mail correspondence.  

Disease severity was assessed at baseline using all four parts of the UPDRS: I (mentation, behaviour, mood), II (activities of daily living, ADL), III (motor function), and IV (complications of therapy). Subtotals I-IV were obtained by adding scores of individual items of the UPDRS in Parts I-IV, respectively. Composite UPDRS scores for tremor (items 20 and 21), rigidity (item 22), bradykinesia (items 23, 25, 26, and 31) and Postural Instability and Gait (PIGD, items 13, 14, 15, 27, 28, 29, 30) were also calculated. In this paper, subtotals and composite UPDRS scores are referred to as aggregate measures.

\subsection{Classification methods}
Four classification methods were evaluated for the identification of UPDRS-related factors associated with falls: decision trees (DTs), random forests (RFs), logistic regression with forward variable selection and logistic regression with Bayesian model averaging (BMA).  In this section, key details of each method and selected criteria for model comparison are outlined.

\begin{figure*}[ht]
%\label{figA4}
\centering
\begin{subfigure}[b]{.45\textwidth}
  \centering
  \includegraphics[keepaspectratio=true,scale=0.45]{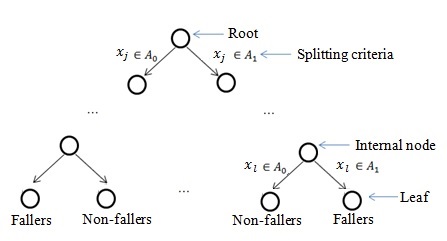}
  \caption{Decision Tree}
  \label{figDT}
\end{subfigure}%
~
\begin{subfigure}[b]{.45\textwidth}
  \centering
  \includegraphics[keepaspectratio=true,scale=0.45]{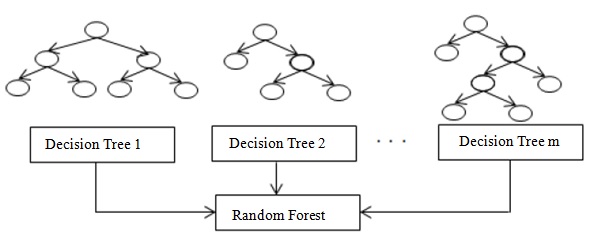}
  \caption{Random Forest}
  \label{figRF}
\end{subfigure}%

~
\begin{subfigure}[b]{.45\textwidth}
  \centering
  \includegraphics[keepaspectratio=true,scale=0.45]{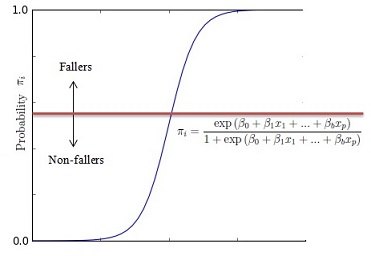}
  \caption{Logistic regression}
  \label{figLOGIT}
\end{subfigure}
~
\begin{subfigure}[b]{.45\textwidth}
  \centering
  \includegraphics[keepaspectratio=true,scale=0.45]{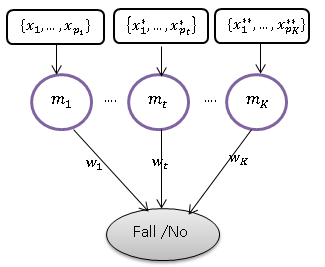}
  \caption{BMA}
  \label{figBMA}
\end{subfigure}
\caption{Classification methods used to classify falls in Parkinson's patients.}
\label{fig:ClassMethods}
\end{figure*}

Decision trees apply recursive partitioning to identify the subset of predictors that best discriminates observations into different categories of the outcome variable (faller/non-faller).  A decision tree procedure begins by determining with predictor best splits the data into two nodes to minimise classification entropy \cite{Breiman2006, Gini1912}.  For continuous predictors, splits are in the form of an optimal cut-off ($\geq$, $\leq$).  Splits on categorical predictors are in the form of membership to a chosen category.  This binary partitioning is then repeated on resulting groups until minimum criteria on node size and/or changes in the chosen misclassification criterion are met. A schematic of this process is provided in Figure \ref{figDT}.
Decision trees results in this paper were obtained using rpart package \cite{Therneau2010a} in R 3.2.5 \cite{RCoreTeam2016}. 

Random forests \cite{Breiman2001}(Figure~\ref{figRF}) offer a robust alternative to decision trees that incorporate bootstrapping and random predictor selection to reduce uncertainty in model predictions. This method fits multiple decision trees, where each tree is fitted to a random subset of the data, sampled with replacement. Within a single tree, splits are determined from a random subset of predictors sampled at each node, as a means of reducing correlation among predictors \cite{Amit1997,Breiman2000}. A consensus prediction for a single observation is obtained by combining predictions across trees; for categorical outcomes, the consensus prediction is the most commonly predicted classification over all trees. Random forests are therefore viewed as a form of model averaging \cite{Bostrom2007}.
Data processing for RF were conducted using randomForest package \cite{Breiman2006} in R 3.2.5 \cite{RCoreTeam2016}.

Turning to the regression based methods, logistic regression can be appropriate when the response variable is dichotomous. The underlying relationships between the explanatory and response variables can be explained by the regression model, through the regression coefficients $\boldsymbol{\beta}=(\beta_0, ..., \beta_K)$ and covariate information $\textbf{x}=(x_1, ..., x_K)$, as
\begin{equation*}
\text{logit} (\pi_{i})=\beta_{0}+\beta_{1}x_{1i}+...+\beta_{K}x_{Ki},
\end{equation*}  
where $\pi_i$ is the probability of fall for patient $i$ having $K$ measurement $\textbf{x}_i=(x_{1i}, ..., x_{Ki})$. 
The relative risk, or odds ratio, of being in one class of the response (i.e. observing a fall) based on a specified value of the explanatory variables, say $x_{j}$, can be predicted by taking the exponentiation of the corresponding regression coefficient, $e^{\beta_{j}}$. Odds ratio greater than one would suggest that the explanatory variable being considered is associated with increase of risk of getting an event, and the opposite is for odds ratio less than one. Odds ratio equal to one simply states non-association of the explanatory and response variables. 

Within a Bayesian framework, one needs to specify the prior distribution for the model parameters. Here, each regression coefficient is assumed to follow a Gaussian distribution, $\beta_j \sim N(0,v_0)$. To represent a vague prior knowledge, $v_0$ is set to a large value (i.e. $10^3$ in R-INLA \cite{Rue2009}).

When fitting a logistic regression model, it is necessary to only include the important explanatory variables in the model.
This problem of variable selection is not adequately addressed in many PD studies mentioned earlier \cite{Allcock2009, Kerr2010, Pickering2007, Sun2009, Wood2002, Gazibara2015,Drapier2005, Fahn2004}. Among the few that raised concern about selection of  important variables is \cite{Kerr2010} that used the total scores (of different clinical instruments) rather than their component (or item) scores to avoid redundancy in the variables used. While, \cite{Hoskovcova2015} included variables that were significant in the univariate model. However, in the presence of other covariates, the contribution of a predictor variable to the prediction could change from the univariate case. Thus this approach does not ascertain that only important explanatory variables are included in the model. 

Determining variables to include in the model is a problem of model choice, and is generally quite a difficult problem to solve in practice due to the large range of potential models. Here, we tackle this problem via a forward variable selection procedure using the log marginal likelihood (hereafter will be denoted by $lml$) to make decisions about whether a variable should be included or excluded from the model. The likelihood is used to measure how the model fits the data \cite{Gelman2014}. Marginalizing it over the set of parameters produces a marginal likelihood, a well established model selection criterion in Bayesian statistics \cite{Hubin2016}. By the marginalizing process, it accounts for all the model's parameters which implies a trivial inbuilt penalty for model complexity. For the numerical stability reason, the marginal likelihood is generally computed in logarithmic scale, resulting an $lml$. It is difficult (in some cases are impossible) to calculate the marginal likelihood analytically as most of the models contain unknown parameters, and thus an approximation is required. Among many approaches to approximate the marginal likelihood, Integrated nested Laplace approximation (INLA) \cite{Rue2009} has become a popular choice for its computationally fast yet still reasonably precise \cite{Hubin2016}. 

In forward variable selection procedure, variables are added to the model one at a time. Starting with a model consisting of an intercept term only,  at each step, each variable that is not in the model is tested for inclusion in the model. The variable that results the highest $lml$ is included in the model, as long as the $lml$ is higher than that of the current model. The process continues until there is no more increase in the $lml$. We denote the model consisting these selected variables as the 'preferred' model.

While the forward variable selection procedure should yield a selection of important variables, it ignores model uncertainty. To overcome this problem we consider several potential models, where the prediction is made upon averaging the results from these models. The procedure starts by taking the logistic regression model identified by forward variable selection, then, all models considered in the variable selection procedure are fitted, and predictions are made. A final prediction, called BMA prediction, is then calculated by the weighted average of predictions from all considered models, with the ratio of the model's $lml$ to the total $lml$ of all models as the weights. The logistic regression results in this paper are produced using R-INLA package \cite{Rue2009} in R 3.2.5 \cite{RCoreTeam2016}.

%%%%%%%%%%%%%%%%%%%%%%%%%
\subsection{Model schemes}

For each classification method, seven different subsets of predictors (representing seven model schemes) were proposed for the prediction of falls status (faller, non-faller) at both six and twelve months follow-up. In each subset, several models varied by the selected variables were fit. A collection of these models are contained in the sets of models as listed in (Table \ref{tab:models}). The 'preferred' model, given by the selected variables producing optimal classification rates, from each set of model is chosen and is used for further analysis. 

UPDRS I to UPDRS IV models were fit and compared in order to assess the relative importance of each of the four parts of the UDPRS, and within each part, to identify the relative importance of items. Whereas important items could be identified from these models, this does not necessarily mean that these items play significant role to predict fall/non-fall in the presence of other items from different (probably more important) parts of the UPDRS. Thus UPDRS model consisting combination of all items was also fit. As to the aggregate measures, similar procedure was done for \textit{Subtotal} and \textit{Composite} models, for assessing the relative importance of two different ways of summarizing information from the UPDRS, through subtotal scores and composite measures. Finally, to infer the optimal way of utilizing the UPDRS in predicting falls, a comparison between UPDRS \textit{Subtotal} and \textit{Composite} models is conducted.
\begin{table*}
\centering
\caption{Subsets of predictors for models fitted at 6-month and 12-month of follow-up.}
\label{tab:models}
\scalebox{0.9}{
\begin{tabular}{@{}llc@{}}\toprule
\rowcolor[rgb]{ .949,  .949,  .949} Sets of models & Predictor variables\\ \midrule
UPDRS I & UPDRS I items \\
UPDRS II & UPDRS II items \\
UPDRS III & UPDRS III items \\
UPDRS IV & UPDRS IV items \\
UPDRS & all UPDRS items \\
Subtotal & subtotals 1-4\\
Composite & tremor, rigidity, bradykinesia, PIGD\\
\bottomrule
\end{tabular}
}
\end{table*}

\subsection{Model assessment}

All models in each classification method were assessed by their ability to predict new data by the leave-one-out cross validation method. The classification rates were in the form of sensitivity (or true positive rate, TPR), specificity (1-false positive rate, FPR), and accuracy, and calculated as follows:

\begin{equation}
sensitivity=\frac{TP}{TP+FN}
\end{equation}
\begin{equation}
specificity=\frac{TN}{TN+FP}
\end{equation}
\begin{equation}
accuracy=\frac{TP+TN}{TP+FP+TN+FN}
\end{equation}
with
\begin{enumerate}[label=(\alph*)]
\item TP, true positives, is the number of patients who actually fell and classified as fallers.
\item FP, false positives, is the number of patients who actually did not fall and classified as fallers.
\item TN, true negatives, is the number of patients who actually did not fall and classified as non-fallers.
\item FN, false negatives, is the number of patients who actually did not fall and classified as non-fallers.
\end{enumerate} 

For the logistic regression, once the predicted probabilities are obtained, the classification is based on a chosen threshold. Options for the classification thresholds ranges from 0 to 1. If the predicted probability is greater than the threshold, then it is classified as a faller, and vice versa. The threshold that was actually used was the value jointly optimize the sensitivity and specificity. 

In addition, ROC curves and the corresponding area under the ROC curves (AUC) were also presented for models assessment. The graph of ROC reflects the accuracy of the diagnostic test. ROC near the diagonal line means the model is not useful, as the prediction is not different than the random guess. A good model fit will produce ROC close to the upper left corner of the graph, where the TPR (sensitivity) is close to 1 and FPR (1-specificity) is close to 0. Graphs of ROC curves in this paper were produced using ROCR \cite{Sing2005} package in R 3.2.5 \cite{RCoreTeam2016}.

\FloatBarrier
\section{Results}
\label{Sec:result}
\let\Oldsubsection\subsection
\renewcommand{\subsection}{\FloatBarrier\Oldsubsection}

\subsection{Participants description} 
%\justify

Table \ref{table.desc} summarizes the subjects classified by fallers (those who experienced at least 1 fall during the follow-up period) and non-fallers (those who did not fall), at 6-month and 12-month of follow-up period. 
\begin{table}[htbp]
  \centering
  \caption{Descriptive statistics for study cohort classified by fallers and non-fallers at 6-month and 12-month of follow-ups. Each categorical variable is summarized by the frequency (\%). Numerical variables are summarized by mean (standard deviation). p-value is the statistical significance for Mann-Whitney-Wilcoxon test (for quantitative variables) and chi-square test (for categorical variable).}
    \label{table.desc}
\scalebox{0.8}{
    \begin{tabular}{llll|lll}
    \toprule
         & \multicolumn{3}{c|}{\textbf{6-month}} & \multicolumn{3}{c}{\textbf{12-month}} \\
\cmidrule{2-7}         & \textbf{Fallers } & \textbf{Non-Fallers } & \textbf{p-value} & \textbf{Fallers } & \textbf{Non-Fallers } & \textbf{p-value} \\
    \midrule
    \midrule
    \multicolumn{4}{l|}{\textbf{Demographic}} &      &      &  \\
    \textbf{Gender} &      &      &      &      &      &  \\
  \quad  Male & 16 (42\%) & 22 (58\%) & 0.21 & 19 (50\%) & 19 (50\%) & 0.45 \\
  \quad  Female & 9 (69\%) & 4 (31\%) &      & 9 (69\%) & 4 (31\%) &  \\
    Age (year) & 65.9 (8.2) & 67.2(7.8) & 0.90  & 67.1 (7.9) & 65.8 (6.9) & 0.70 \\
    \multicolumn{4}{l|}{\textbf{Living arrangement}} &      &      &  \\
       \quad Alone & 3 (60\%) & 2 (40\%) & 0.90  & 3 (60\%) & 2 (40\%) & 0.98 \\
       \quad With family & 22 (48\%) & 24 (52\%) &      & 25 (54\%) & 21 (46\%) &  \\
         &      &      &      &      &      &  \\
    \multicolumn{4}{l|}{\textbf{Disease specific}} &      &      &  \\
    UPDRS scores &      &      &      &      &      &  \\
     \quad   Subtotal 1 & 2.9 (2.5) & 1.9 (1.7) & 0.20  & 2.9 (2.5) & 1.8 (1.7) & 0.15 \\
\quad        Subtotal 2 & 14.4 (6.1) & 9.3 (3.9) & \textbf{0.03} & 12.8 (5.9) & 9.6 (3.9) & 0.13 \\
     \quad   Subtotal 3 & 21.8 (9.4) & 15.1 (8.9) & \textbf{0.01} & 20.5 (10.2) & 16.2 (9.1) & \textbf{0.09} \\
        \quad Subtotal 4 & 2.7 (2.3) & 1.3 (1.8) & \textbf{0.08} & 2.5 (2.5) & 1.2 (1.5) & \textbf{0.05} \\
        \quad Total & 41.5 (13.7) & 27.2 (12.3) & \textbf{0.00} & 38.7 (14.8) & 28.7 (12.1) & \textbf{0.03} \\
    Tremor & 3.0 (4.2) & 3.1 (2.2) & 0.20  & 2.6 (3.9) & 3.3 (2.2) & \textbf{0.06} \\
    Rigidity & 3.52 (3.1) & 3.1 (3.3) & 0.40  & 3.6 (2.9) & 3.3 (3.5) & 0.47 \\
    PIGD & 4.9 (2.9) & 3.1 (2.5) & \textbf{0.05} & 4.3 (2.9) & 3.5 (2.5) & 0.37 \\
    Bradykinesia & 7.7 (4.1) & 4 (3.3) & \textbf{0.01} & 7.2 (4.3) & 4.2 (3.5) & \textbf{0.01} \\
    Duration & 6.7 (4.8) & 4.6 (2.7) & 0.40  & 6.7 (4.2) & 4.4 (2.4) & \textbf{0.08} \\
    Previous falls & 1.7 (1.6) & 0.5 (1.1) & \textbf{0.01} & 1.4 (1.5) & .5 (1.3 ) & \textbf{0.02} \\
    \bottomrule
    \end{tabular}%
}
  \label{tab:addlabel}%
\end{table}%

The majority of the participants were males. 
Proportion of fallers and non-fallers is around the same for males, while for females, proportion of fallers is around twice of non-fallers. However, this difference is not statistically significant, as implied by the chi-squared independence test. Similarly, there is no significant difference between age, on the average, between the two groups. Although the number is small, the relative proportion of fallers was higher in people who lived alone than in those who lived with family. Yet, the difference is not significant. Overall, there are no significant differences between fallers and non-fallers based on their demographic information. 

As for the disease specific measurements, UPDRS sub-totals and total scores (except for Subtotal 1 at 6 month and Subtotal 1 and Subtotal 2 at 12 month of follow up) discriminate the two groups. Bradykinesia shows consistency of discriminating fall/non-fall groups at the two follow-up times, while rigidity shows the opposite. Prospective fallers had been diagnosed with the disease for a slightly longer time and had more falls prior to the participation in the study, compared to the non-fallers. 

\subsection{Univariate logistic regression models.} 

\begin{figure}[htp]
\centering\includegraphics[width=.7\linewidth]{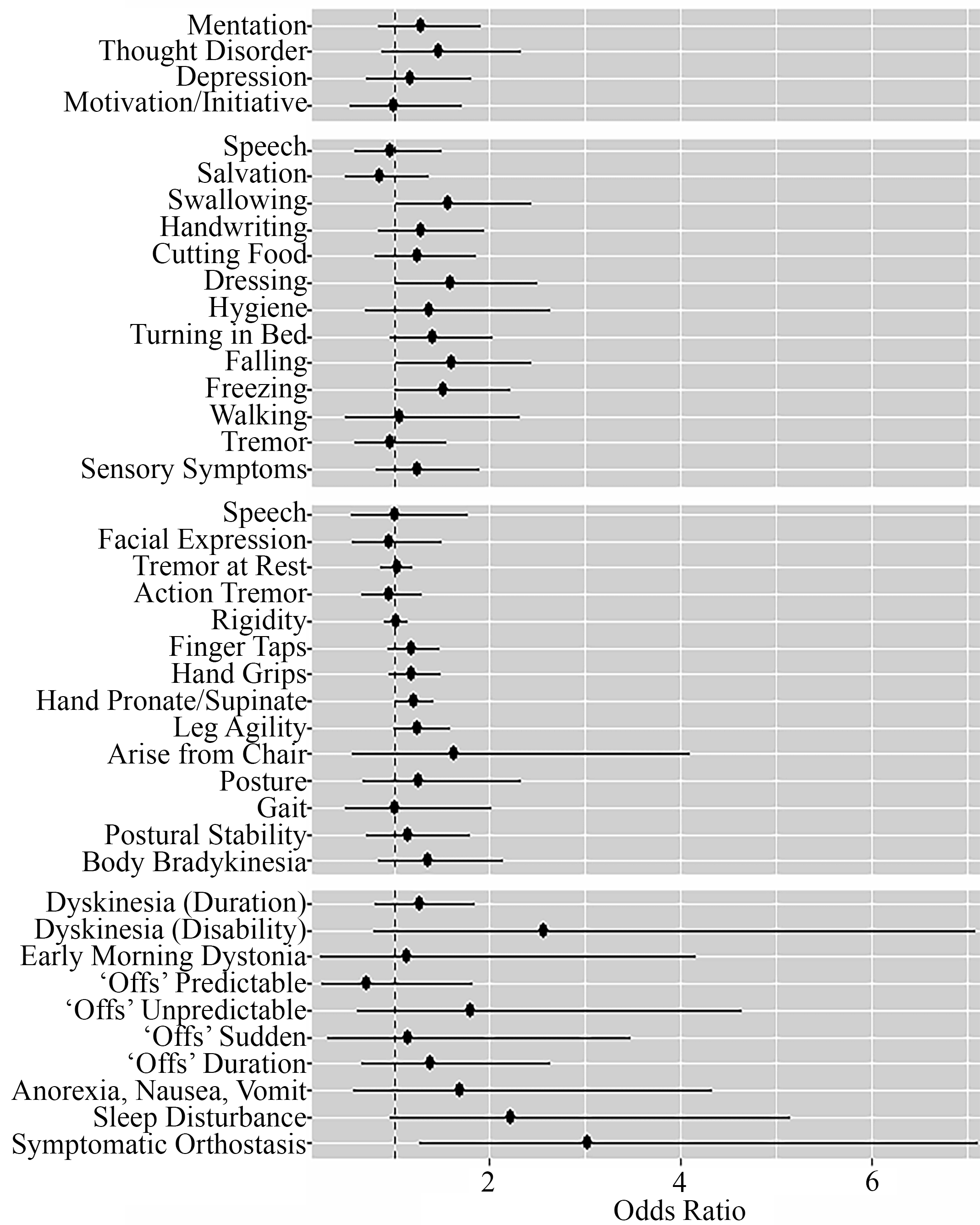}
\caption{Odds ratio (with 95\% CI) of falls classification at 6-month of follow-up, using univariate logistic regression with individual items of the UPDRS as the predictor.}
\label{fig:or items}
\end{figure}

Further exploratory on individual items of the UPDRS were conducted through fitting univariate logistic regression model. Each item is used as an explanatory for fall/non-fall prediction. The results were summarized in the form of an odds ratio (OR, with 95\% confidence intervals) of falling (at 6-month) given the item measurements (Figure~\ref{fig:or items}). Several items produced non-unity ORs: swallowing, dressing, falling (unrelated to freezing) and freezing for UPDRS II, and hand pronate/supinate and leg agility for UPDRS III. This indicates the usefulness of these items in explaining falls. 

\begin{figure}[h]
\centering\includegraphics[width=.6\linewidth]{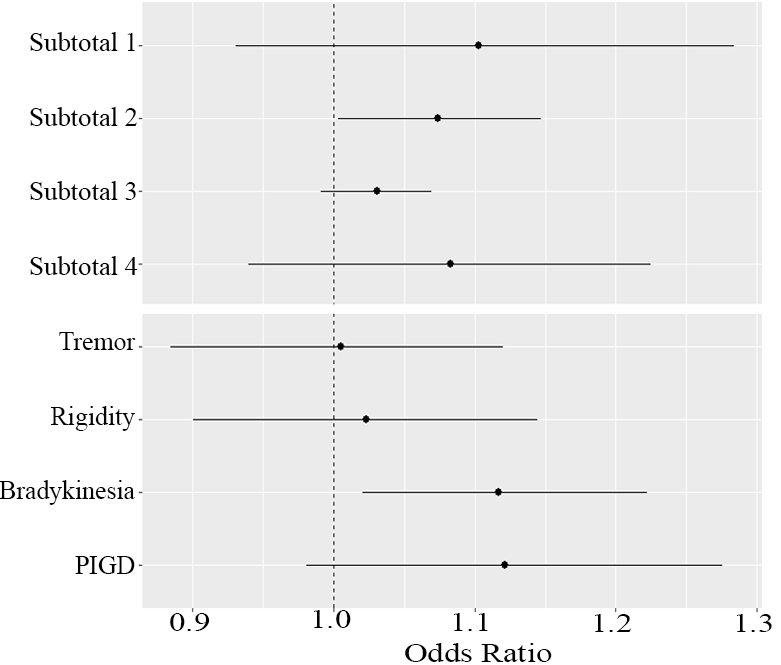}
\caption{Odds ratio (with 95\% CI) of falls classification at 6-month of follow-up, using univariate logistic regression model with aggregate measures of the UPDRS as the explanatory variable. Subtotals 1-4 are the sums of item scores in UPDRS Parts I - IV, respectively.}
\label{fig:or composites}
\end{figure}

Similar exploratory were also conducted for the aggregate measures, and is shown in Figure~\ref{fig:or composites}. Subtotal 2 and bradykinesia produced ORs greater than 1, indicating higher risk of falls for patients with higher score of these aggregate measures. Graphs for univariate tests at 12-month are given in \ref{appendixOR}. 
 
Overall, the plots suggest the usefulness of UPDRS individual items, as well as the composite measures, for falls prediction. However, this result should be used just as such motivate a modelling approach. The condition of people with PD is affected by interdependent factors, some were measured by the UPDRS, and thus the associations between items and falls occurrences might change when other measures are taken into account. Thus, a multivariate model is preferred rather than univariate analysis.

\subsection{Relative importance of UPDRS Parts I - IV}
\begin{figure*}[t!]
%\label{figA4}
\centering
\begin{subfigure}[b]{.5\textwidth}
  \centering
  \includegraphics[keepaspectratio=true,scale=0.4]{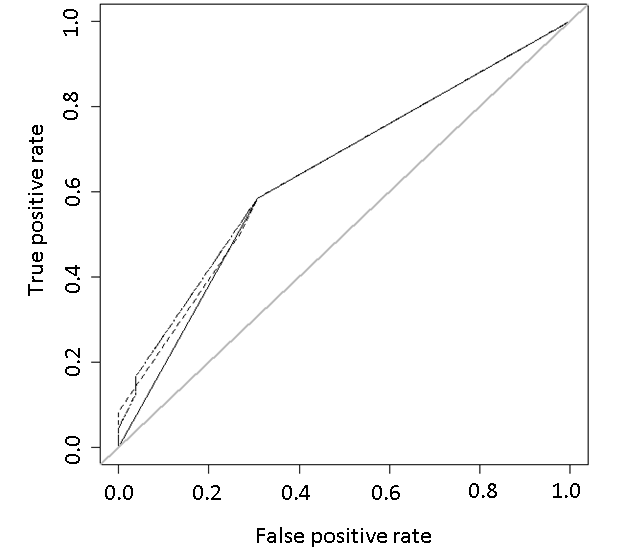}
  \caption{UPDRS I model.}
  \label{roc1}
\end{subfigure}%
~
\begin{subfigure}[b]{.5\textwidth}
  \centering
  \includegraphics[keepaspectratio=true,scale=0.4]{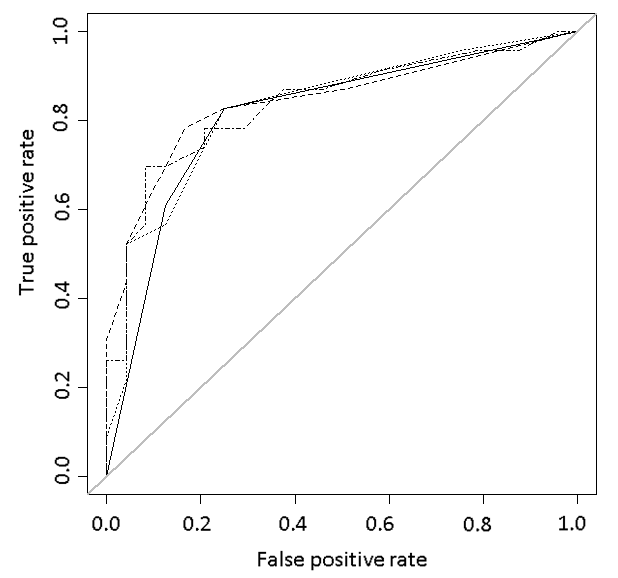}
  \caption{UPDRS II model.}
  \label{roc2}
\end{subfigure}%
~

\begin{subfigure}[b]{.5\textwidth}
  \centering
  \includegraphics[keepaspectratio=true,scale=0.4]{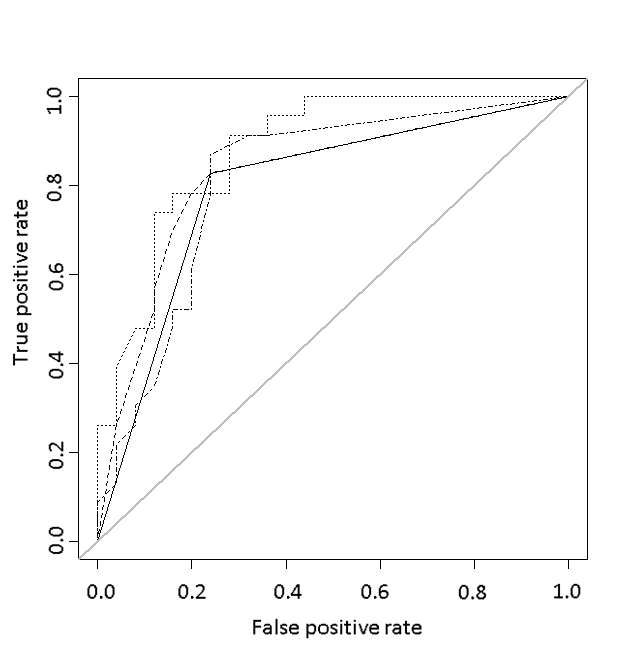}
  \caption{UPDRS III model.}
  \label{roc3}
\end{subfigure}%
~
\begin{subfigure}[b]{.5\textwidth}
  \centering
  \includegraphics[keepaspectratio=true,scale=0.4]{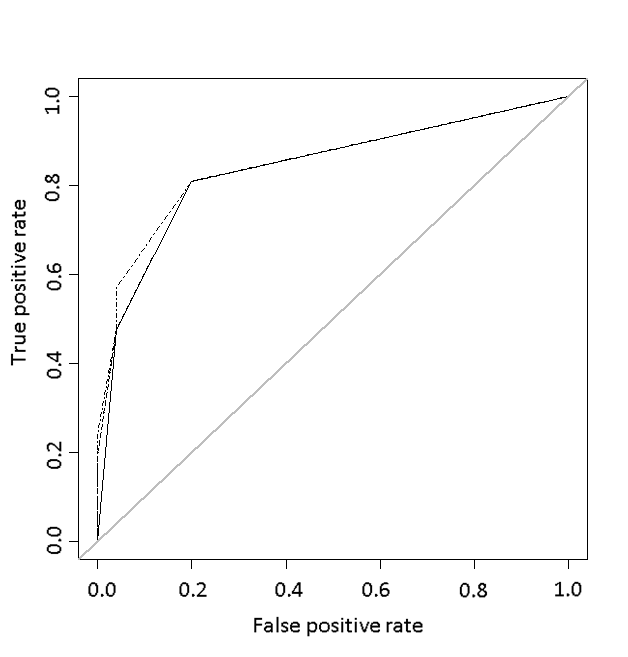}
  \caption{UPDRS IV model.}
  \label{roc4}
\end{subfigure}
\caption{ROC curves for falls classification at 6 month of follow-up using items of each part of the UPDRS. Classification methods employed are: Decision Tree (solid), Random Forest (dashed), logistic regression with stepwise (dotted), and logistic regression with BMA (dashed dotted).}
\label{fig:rocpart}
\end{figure*}

%\justify
For the  interpretation, the results of logistic regression from a forward variable selection procedure will be referred to as LOGIT, and the results from model averaging procedure will be referred to as BMA. 
Table \ref{acc4models} presents classification rates (accuracy, sensitivity, and specificity) and AUC for UPDRS I - IV models. ROC curves are depicted in Figure \ref{fig:rocpart}. In general, all methods agree that items of UPDRS II and UPDRS III can classify participants into fall/non-fall groups better than UPDRS I or UPDRS IV, as these two models produce high accuracy, sensitivity and specificity, at 6- and 12-month follow-up times. This is also confirmed by the high values of AUC for UPDRS II and III models. Between UPDRS I and IV, items of the latter part are more informative than UPDRS I items, as implied by higher classification rates for UPDRS IV model than that for UPDRS I model.

\begin{table}[htp]
  \centering
  \caption{Classification rates (accuracy, sensitivity, specificity) and AUC of models with individual items of UPDRS Part I-IV as the explanatory variables. Highest values among the four models within each method are in bold. LOGIT is the logistic regression with forward variable selection, and BMA is the logistic regression with the Bayesian model averaging.}
	\label{acc4models}
\scalebox{0.55}{\begin{tabular}{l|cccc|cccc|cccc|cccc}
    \multicolumn{17}{c}{(a) At 6-month of follow-up} \\
    \midrule
    \rowcolor[rgb]{ .949,  .949,  .949} \multicolumn{1}{c|}{Model} & \multicolumn{4}{c|}{Accuracy} & \multicolumn{4}{c|}{Sensitivity} & \multicolumn{4}{c|}{Specificity} & \multicolumn{4}{c}{AUC} \\
\cmidrule{2-17}    \rowcolor[rgb]{ .949,  .949,  .949}      & DT   & RF   & LOGIT & BMA  & DT   & RF   & LOGIT & BMA  & DT   & RF   & LOGIT & BMA  & DT   & RF   & LOGIT & BMA \\
    \midrule
    \midrule
    UPDRS I & 0.52 & 0.52 & 0.52 & 0.52 & 0.47 & 0.47 & 0.47 & 0.47 & 0.56 & 0.56 & 0.56 & 0.56 & 0.52 & 0.53 & 0.53 & 0.53 \\
    UPDRS II & \textbf{0.71} & 0.70 & \textbf{0.75} & 0.71 & \textbf{0.75} & \textbf{0.75} & 0.70 & 0.70 & \textbf{0.68} & \textbf{0.73} & \textbf{0.79} & \textbf{0.71} & \textbf{0.73} & \textbf{0.76} & \textbf{0.77} & \textbf{0.76} \\
    UPDRS III & \textbf{0.71} & \textbf{0.71} & 0.71 & \textbf{0.73} & \textbf{0.75} & \textbf{0.75} & \textbf{0.82} & \textbf{0.78} & \textbf{0.68} & 0.68 & 0.61 & 0.68 & 0.71 & 0.74 & 0.73 & 0.73 \\
    UPDRS IV & 0.65 & 0.65 & 0.65 & 0.65 & 0.66 & 0.66 & 0.66 & 0.65 & 0.65 & 0.65 & 0.65 & 0.65 & 0.67 & 0.68 & 0.68 & 0.68 \\
    \midrule
    \multicolumn{1}{r}{} &      &      &      & \multicolumn{1}{c}{} &      &      &      & \multicolumn{1}{c}{} &      &      &      & \multicolumn{1}{c}{} &      &      &      &  \\
    \multicolumn{17}{c}{(b) At 12-month of follow-up} \\
    \midrule
    \rowcolor[rgb]{ .949,  .949,  .949} \multicolumn{1}{c|}{Model} & \multicolumn{4}{c|}{Accuracy} & \multicolumn{4}{c|}{Sensitivity} & \multicolumn{4}{c|}{Specificity} & \multicolumn{4}{c}{AUC} \\
\cmidrule{2-17}    \rowcolor[rgb]{ .949,  .949,  .949}      & DT   & RF   & LOGIT & BMA  & DT   & RF   & LOGIT & BMA  & DT   & RF   & LOGIT & BMA  & DT   & RF   & LOGIT & BMA \\
    \midrule
    \midrule
    UPDRS I & 0.50 & 0.57 & 0.50 & 0.50 & 0.45 & 0.48 & 0.45 & 0.45 & 0.57 & 0.67 & 0.57 & 0.57 & 0.51 & 0.64 & 0.53 & 0.53 \\
    UPDRS II & \textbf{0.73} & 0.73 & \textbf{0.73} & \textbf{0.75} & 0.73 & 0.73 & 0.69 & 0.69 & \textbf{0.73} & \textbf{0.73} & \textbf{0.77} & \textbf{0.81} & \textbf{0.75} & 0.77 & \textbf{0.77} & \textbf{0.77} \\
    UPDRS III & 0.71 & \textbf{0.79} & 0.71 & 0.71 & \textbf{0.83} & \textbf{0.83} & \textbf{0.79} & \textbf{0.73} & 0.58 & 0.74 & 0.61 & 0.69 & 0.73 & \textbf{0.84} & 0.71 & 0.71 \\
    UPDRS IV & 0.63 & 0.63 & 0.65 & 0.65 & 0.61 & 0.61 & 0.64 & 0.64 & 0.66 & 0.66 & 0.66 & 0.66 & 0.65 & 0.66 & 0.68 & 0.69 \\
    \bottomrule
    \end{tabular}%
}
  \label{tab:addlabel}%
\end{table}%

More insight into UPDRS II and UPDRS III models at 6-month of follow-up, the accuracy is not significantly different for these two models. All the methods produce almost the same classification rates, with the range between 71\% to 75\%. DT and RF yield the same sensitivity at 75\%, while LOGIT and BMA produce higher sensitivity for UPDRS III model, at 82\% and 78\% respectively. The opposite is observed for the specificity, with the higher values are for UPDRS II model, at 71\% to 79\% as produced by RF, LOGIT, and BMA. Similar with accuracy, the difference in AUC is negligible (only 2\%) -as also implied by overlap ROC curves in Figures \ref{fig:rocpart}(b) and \ref{fig:rocpart}(c)- thus confirms the comparable results of UPDRS II and UPDRS III models. 

Similar trend is also obtained for the 12-month follow-up. The accuracy of UPDRS II and UPDRS III models are between 73\% to 79\%, with the highest is 79\% for UPDRS III model using RF. All methods consistently produce higher sensitivity for UPDRS III model than that for UPDRS II model, with the highest at 83\% produced by DT and RF. While, the specificity for UPDRS II is consistently higher (for the 4 implemented methods) than that of UPDRS III, with the highest specificity at 81\% produced by BMA. The difference in AUC for UPDRS II and UPDRS III models is only 2\% at 6-month period, and more varied (from 2 to 7\%) at 12-month period. However, AUC does not clearly differentiate the two models. Despite higher AUC for UPDRS II model produced by 3 methods other than RF, the highest AUC at 0.84 is obtained from UPDRS III model using RF.  

%%%%%%%%%%%%%%%%%%%%%%%%
\newcommand\solidrule[1][1cm]{\rule[0.5ex]{#1}{.4pt}}
\newcommand\dashedrule{\mbox{%
  \solidrule[2mm]\hspace{2mm}\solidrule[2mm]\hspace{2mm}\solidrule[2mm]}}
%%%%%%%%%%%%%%%%%%%%%%%%
	
Comparing the two follow-up times, the differences in accuracy at 6-month and 12-month are not significant for all models. There are variations in sensitivity, specificity, and AUC values produced by the same method for the same model. For example, sensitivity for UPDRS III model is higher at 12-month than that at 6-month based on DT and RF, and the opposite is for UPDRS II model. However, LOGIT and BMA produced lower sensitivity at 12-month than that at 6-month for these models. Yet, the relatively small differences can be neglected. Similar variations also obtained for specificity and AUC. Thus, it can be inferred that 6-month and 12-month time periods produce comparably similar results with regards to falls prediction based on items of UPDRS parts I - IV.    

Comparing the methods, all produced relatively the same results, with slight variations for UPDRS II and UPDRS III models. The accuracy is slightly higher -on average- for LOGIT and BMA in UPDRS II and UPDRS III models at 6-month and UPDRS II model at 12-month period. On the other hand, DT and RF produce a slightly higher sensitivity than LOGIT and BMA, as can be seen in UPDRS II model at 6-month and UPDRS II and UPDRS III models at 12-month. The specificity from DT and RF are less than that from LOGIT and BMA only -on average- for UPDRS II model at 12-month period. However, all the differences from these comparison are relatively small, less than 5\% on average. Thus, it can be inferred that among the 4 methods implemented for this study, none is significantly produce different results.     

Overall, individual items of parts II and III of the UPDRS are useful explanatory for falls classification. It cannot be decided clearly which of these two parts is more accurate to predict fall/non-fall, as the differences in accuracy of UPDRS II model is not distinct from that of UPDRS III model. However, further checking indicates that UPDRS III items tend to be more sensitive than UPDRS II items in identifying fallers for their higher sensitivity. While if the focus is in identifying non-faller, then UPDRS II model is preferred as its specificity is higher than that of UPDRS III model.

%%%%%%%%%%%%%%%%%%%%%%%
\subsection{Predictive comparison between the individual items and summary quantities of UPDRS}
%\justify

\begin{figure*}[t!]
%\label{figA4}
\centering
\begin{subfigure}[b]{.3\textwidth}
  \centering
  \includegraphics[keepaspectratio=true,scale=0.25]{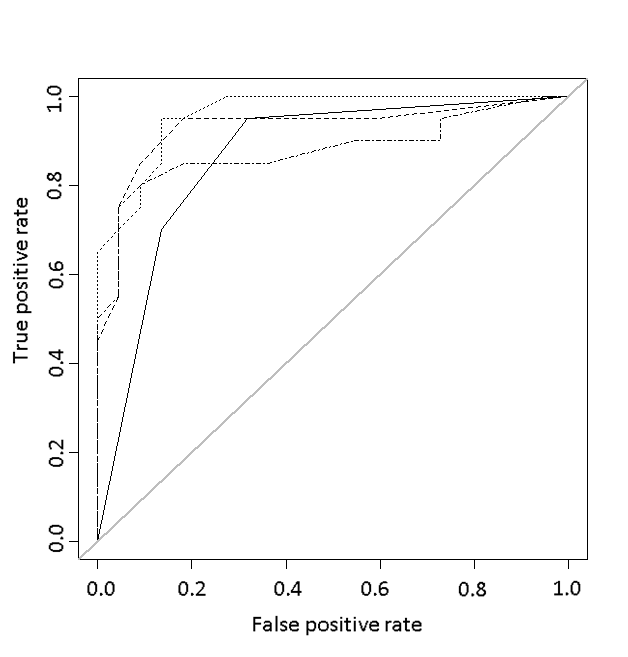}
  \caption{UPDRS model.}
  \label{roc5}
\end{subfigure}%
~
\begin{subfigure}[b]{.3\textwidth}
  \centering
  \includegraphics[keepaspectratio=true,scale=0.26]{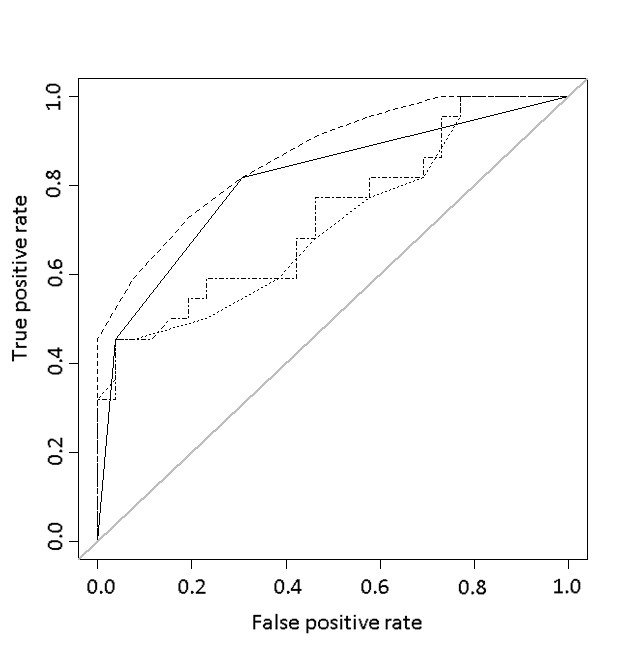}
  \caption{\textit{Subtotal} model.}
  \label{roc6}
\end{subfigure}%
~
\begin{subfigure}[b]{.3\textwidth}
  \centering
  \includegraphics[keepaspectratio=true,scale=0.26]{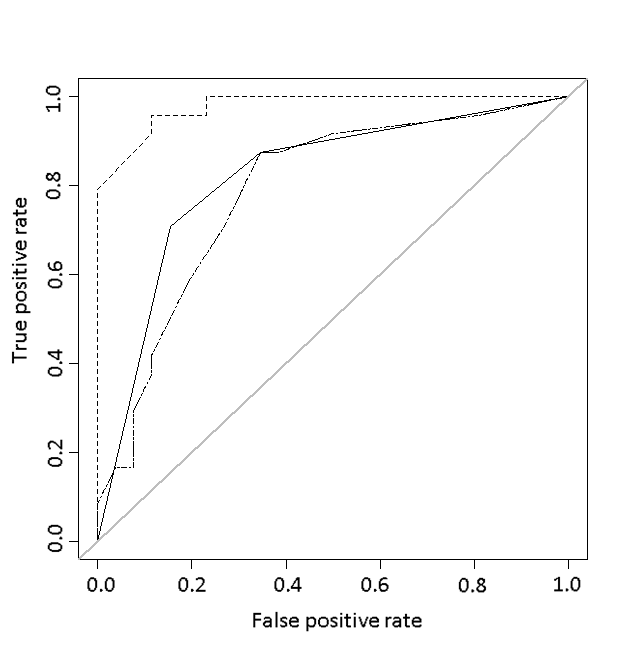}
  \caption{\textit{Composite} model.}
  \label{roc7}
\end{subfigure}%
\caption{ROC curves for falls classification at 6-month of follow-up using individual items (a), composite measures (b) and individual items and composite measures of the UPDRS. Classification methods employed are: Decision Tree (solid), Random Forest (dashed), logistic regression with stepwise (dotted), and logistic regression with BMA (dashed dotted).}
\label{fig:roc.composite}
\end{figure*}

The classification rates for UPDRS items, \textit{Subtotal}, and \textit{Composite} models are presented in Table~\ref{acc.composites}, and the corresponding ROC curves are depicted in Figure \ref{fig:roc.composite}. 

\begin{table}[htbp]
  \centering
  \caption{Classification rates (accuracy, sensitivity, specificity) and AUC of models with individual items and aggregate measures of UPDRS as the explanatory variables. Highest values among the four models within each method are in bold. LOGIT is the logistic regression with forward variable selection, and BMA is the logistic regression with the Bayesian model averaging.}
	\label{acc.composites}
	\scalebox{0.55}{
    \begin{tabular}{l|cccc|cccc|cccc|cccc}
    \multicolumn{17}{c}{(a) At 6-month of follow-up} \\
    \midrule
    \rowcolor[rgb]{ .949,  .949,  .949} \multicolumn{1}{c|}{Model} & \multicolumn{4}{c|}{Accuracy} & \multicolumn{4}{c|}{Sensitivity} & \multicolumn{4}{c|}{Specificity} & \multicolumn{4}{c}{AUC} \\
\cmidrule{2-17}    \rowcolor[rgb]{ .949,  .949,  .949}      & DT   & RF   & LOGIT & BMA  & DT   & RF   & LOGIT & BMA  & DT   & RF   & LOGIT & BMA  & DT   & RF   & LOGIT & BMA \\
    \midrule
    \midrule
    UPDRS & \textbf{0.72} & 0.78 & \textbf{0.80} & \textbf{0.78} & \textbf{0.81} & \textbf{0.85} & \textbf{0.85} & \textbf{0.80} & 0.61 & 0.74 & 0.77 & \textbf{0.82} & \textbf{0.77} & 0.83 & \textbf{0.85} & \textbf{0.77} \\
    Subtotal & 0.67 & \textbf{0.84} & 0.65 & 0.61 & 0.73 & 0.82 & 0.40 & 0.49 & 0.61 & \textbf{0.85} & \textbf{0.85} & 0.72 & 0.72 & 0.77 & 0.63 & 0.65 \\
    Composite & 0.69 & 0.80 & 0.68 & 0.68 & 0.63 & 0.78 & 0.78 & 0.78 & \textbf{0.76} & 0.82 & 0.58 & 0.58 & 0.73 & \textbf{0.87} & 0.70 & 0.70 \\
    \midrule
    \multicolumn{1}{r}{} &      &      &      & \multicolumn{1}{c}{} &      &      &      & \multicolumn{1}{c}{} &      &      &      & \multicolumn{1}{c}{} &      &      &      &  \\
    \multicolumn{17}{c}{(b) At 12-month of follow-up} \\
    \midrule
    \rowcolor[rgb]{ .949,  .949,  .949} \multicolumn{1}{c|}{Model} & \multicolumn{4}{c|}{Accuracy} & \multicolumn{4}{c|}{Sensitivity} & \multicolumn{4}{c|}{Specificity} & \multicolumn{4}{c}{AUC} \\
\cmidrule{2-17}    \rowcolor[rgb]{ .949,  .949,  .949}      & DT   & RF   & LOGIT & BMA  & DT   & RF   & LOGIT & BMA  & DT   & RF   & LOGIT & BMA  & DT   & RF   & LOGIT & BMA \\
    \midrule
    \midrule
    UPDRS & \textbf{0.74} & 0.79 & \textbf{0.76} & \textbf{0.79} & 0.83 & 0.83 & \textbf{0.80} & \textbf{0.84} & \textbf{0.71} & 0.76 & \textbf{0.74} & \textbf{0.78} & \textbf{0.77} & \textbf{0.84} & \textbf{0.79} & \textbf{0.81} \\
    Subtotal & 0.60 & \textbf{0.83} & 0.49 & 0.49 & 0.61 & 0.82 & 0.51 & 0.51 & 0.60 & \textbf{0.85} & 0.48 & 0.48 & 0.64 & 0.82 & 0.55 & 0.55 \\
    Composite & 0.68 & 0.77 & 0.63 & 0.63 & \textbf{0.85} & \textbf{0.85} & 0.69 & 0.69 & 0.48 & 0.66 & 0.55 & 0.55 & 0.67 & 0.83 & 0.63 & 0.63 \\
    \bottomrule
    \end{tabular}%
}
  \label{tab:addlabel}%
\end{table}%

In general, UPDRS items are more informative than the aggregate measures as the accuracy and sensitivity of UPDRS model are higher than that of Subtotal and Composite models. An exception is the accuracy produced by RF, which yield higher values for \textit{Subtotal} model, 84\% at 6-month and 83\% at 12-month periods. The sensitivity of UPDRS items model range from 80\% to 85\%, almost double the sensitivity of Subtotal model for LOGIT and BMA at 6-month period. Although lower than that of \textit{Subtotal} model, the specificity of UPDRS items model are still reasonably high, at 74\% to 82\% (except for DT is 61\% at 6-month period). The relatively higher AUC values -as confirmed by ROC curves in Figure \ref{fig:roc.composite}(a) is more to the upper left corner than ROC curves in Figure \ref{fig:roc.composite}(b),(c)- also support that UPDRS items outperform UPDRS subtotals and composite measures to predict fall/non-fall. 

Further comparison of the aggregate measures, at 6-month period, the difference in accuracy between subtotal scores and composite measures is relatively small. However, based on LOGIT and BMA, the sensitivity of \textit{Subtotal} model is far less than that of \textit{Composite} model and the opposite is for the specificity. While of based on DT and RF, the sensitivity is higher for Subtotal model than for Composite model. Furthermore, the difference between sensitivity and specificity is not as large as that based on LOGIT and BMA. The AUC also suggests similar results for \textit{Subtotal} and \textit{Composite} models.

While at 12-month period, despite the relatively low values, the accuracy of \textit{Composite} model is substantially higher by 14\% than \textit{Subtotal} model (based on LOGIT and BMA). Its sensitivity, specificity, and AUC are also higher than the \textit{Subtotal} model. This implies that based on LOGIT and BMA, the composite measures are relatively more informative than the subtotals in predicting fall/non-fall. However, composite model has higher difference between sensitivity and specificity than the \textit{Subtotal} model, and the sensitivity is higher than the specificity. This implies that making a decision as to whether a patient will fall or not is easier using the composite measures than the subtotals. On the other hand, DT and RF yield higher classification rates -on average- for these two models. 

In summary, UPDRS items are shown to outperform the aggregate measures in predicting fall/non-fall. As for the aggregate measures, the difference in classification rate between \textit{Subtotal} model and \textit{Composite} model is relatively small. Composite measures are more sensitive in identifying fallers than identifying non-fallers, compared to the subtotals. As for the methods, the tree-based methods (DT and RF) provide higher classification rates than the regression-based methods (LOGIT and BMA) when aggregate measures are used instead of the UPDRS items.

%%%%%%%%%%%%

\subsection{Important risk factors related to falls}
\label{sec:riskfactors}

Important explanatory variables extracted from the models for DT, RF, and logistic regression with forward variable selection (LOGIT) are listed in Table \ref{important.vars} for the two follow-up times. Variables in BMA for the selected models are listed in \ref{sec:BMAresult}. In general, the four methods selected similar set of variables as shown by several common important variables in each model. 

At 6-month of follow-up, by modelling each of the four parts of the UPDRS separately several items were consistently selected: thought disorder in UPDRS I, dressing and falling in UPDRS II, hand pronate/supinate in UPDRS III, and symptomatic orthostasis and sleep disturbance in UPDRS IV. In addition for LOGIT, freezing was also a significant item in UPDRS Part II, and leg agility in UPDRS Part III. Similar set of items were also selected at the 12-month period, indicating the consistency of the models and methods performance in the two follow-up times.  

When all items were combined together, as in all UPDRS items model, dressing and hand pronate/supinate were always selected by all classification methods, in both time periods. In addition, speech, sleep disturbance, and symptomatic orthostasis were also regarded important in regression.

\begin{figure*}[htbp]
%\label{figA4}
\centering
\begin{subfigure}[b]{.5\textwidth}
  \centering
  \includegraphics[keepaspectratio=true,scale=0.4]{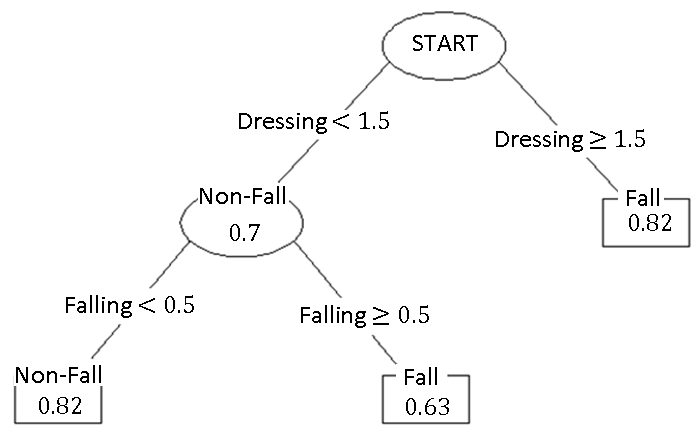}
  \caption{UPDRS II model}
  \label{fig:DT_M2_6m}
\end{subfigure}%
~
\begin{subfigure}[b]{.5\textwidth}
  \centering
  \includegraphics[keepaspectratio=true,scale=0.4]{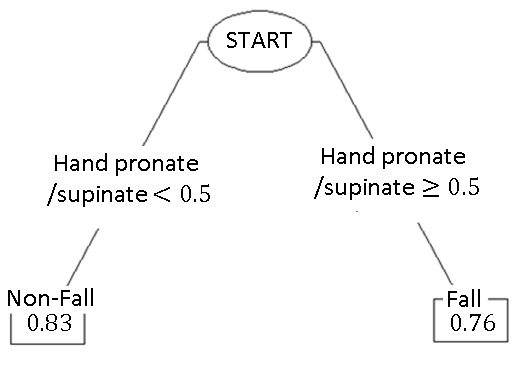}
  \caption{UPDRS III model}
  \label{fig:DT_M3_6m}
\end{subfigure}%
~

\begin{subfigure}[b]{.51\textwidth}
  \centering
  \includegraphics[keepaspectratio=true,scale=0.4]{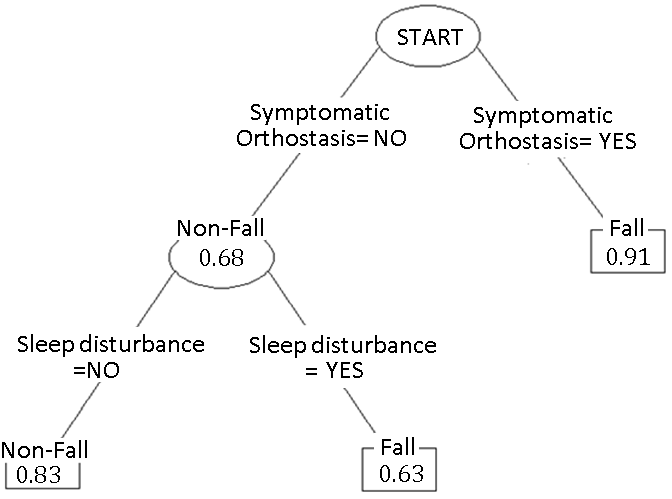}
  \caption{UPDRS IV model}
  \label{fig:DT_M4_6m}
\end{subfigure}%
~
\begin{subfigure}[b]{.5\textwidth}
  \centering
  \includegraphics[keepaspectratio=true,scale=0.4]{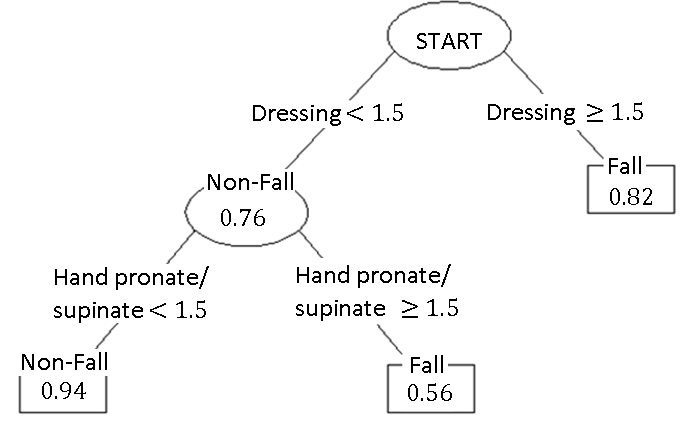}
  \caption{All UPDRS items model}
  \label{fig:DT_M5_6m}
\end{subfigure}%
\caption{Falls prediction using Decision Tree with items of the UPDRS as the predictor variables. Items scoring are based on UPDRS questionnaire scoring. Values in the final nodes are the purity (homogeneity) of the nodes (proportion of correctly classified patients relative to all patients in that node).}
\label{fig:DT_6m_items}
\end{figure*}
%%%%%%%%%%%%
\begin{figure*}[t!]
%\label{figA4}
\centering
\begin{subfigure}[b]{.5\textwidth}
  \centering
  \includegraphics[keepaspectratio=true,scale=0.4]{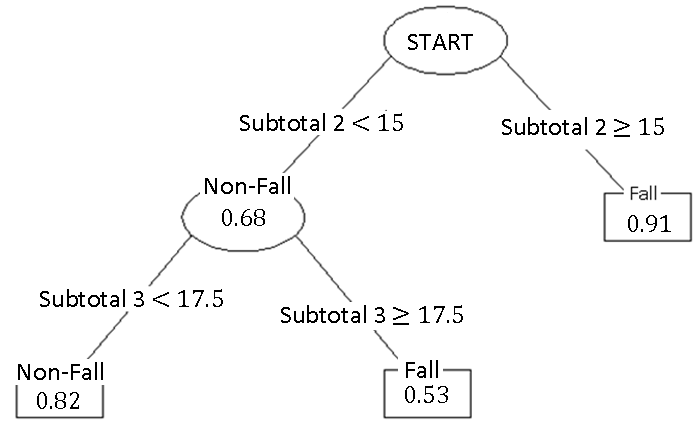} 
  \caption{\textit{Subtotal} model}
  \label{fig:DT_M4_6m}
\end{subfigure}%
~
\begin{subfigure}[b]{.5\textwidth}
  \centering
  \includegraphics[keepaspectratio=true,scale=0.4]{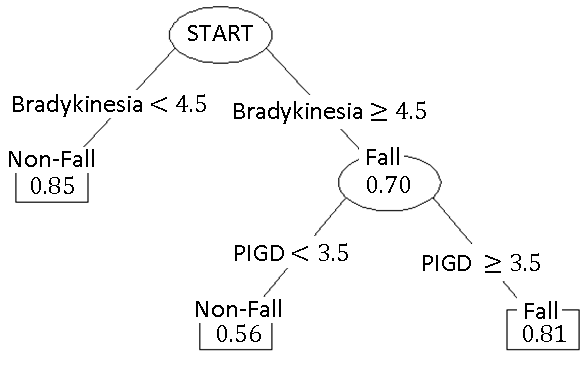}
  \caption{\textit{Composite} model}
  \label{fig:DT_M5_6m}
\end{subfigure}%
\caption{Falls prediction using Decision Tree with aggregate measures of the UPDRS as the predictor variables. Items scoring are based on UPDRS questionnaire scoring. Values in the final nodes are the purity (homogeneity) of the nodes (proportion of correctly classified patients relative to all patients in that node).}
\label{fig:DT_6m_aggregate}
\end{figure*}

As for the aggregate measures, Subtotal 2 and bradykinesia are selected by all the methods at 6-month and 12-month periods. In addition, subtotal 3 was also considered important at 6-month but not at 12-month if the models were based on DT and LOGIT. It it worth to note that at 6-month period, PIGD is selected in the model but is replaced by tremor at 12-month period. 

Among the 4 methods used, DT is appealing for its ease of visualization and interpretation. Figure \ref{fig:DT_6m_items} and Figure \ref{fig:DT_6m_aggregate} exemplify falls prediction at 6-month period using for UPDRS items and aggregate measures as the predictor variables. Prediction using UPDRS I model is not displayed, as it has only 1 predictor variable, thought disorder, and the classification rates are the lowest amongst all models. Through Figure \ref{fig:DT_6m_items} we can infer that examination on several selected (targeted) items could provide information on deciding whether a patient will likely to fall or not-fall based on the cut-offs given from the model without the need to calculate some derived scores such as odds ratios as in logistic regression.

\begin{landscape}
\begin{table}[htbp]
  \centering
  \caption{Selected variables for falls prediction using the classification methods: Decision Tree (DT), Random Forest (RF), and  logistic regression with forward variable selection (LOGIT). }
\label{important.vars}
	\scalebox{0.6}{
    \begin{tabular}{rrrrrlrr}
    \multicolumn{8}{c}{(a) At 6-month period} \\
    \midrule
    \rowcolor[rgb]{ .949,  .949,  .949} \multicolumn{1}{c}{\textbf{Method}} & \multicolumn{1}{c}{\textbf{UPDRS I }} & \multicolumn{1}{c}{\textbf{UPDRS II}} & \multicolumn{1}{c}{\textbf{UPDRS III}} & \multicolumn{1}{c}{\textbf{UPDRS IV}} & \multicolumn{1}{c}{\textbf{All UPDRS items}} & \multicolumn{1}{c}{\textbf{UPDRS subtotals}} & \multicolumn{1}{c}{\textbf{Composite measures}} \\
    \midrule
    \midrule
    \multicolumn{1}{l}{\textbf{DT}} & \multicolumn{1}{l}{Thought disorder} & \multicolumn{1}{l}{Dressing} & \multicolumn{1}{l}{Hand pronate/supinate} & \multicolumn{1}{l}{Sleep disturbance} & Dressing & \multicolumn{1}{l}{Subtotal 2} & \multicolumn{1}{l}{Bradykinesia} \\
         &      & \multicolumn{1}{l}{Falling} &      & \multicolumn{1}{l}{Symptomatic orthostasis} & Hand pronate/supinate & \multicolumn{1}{l}{Subtotal 3} & \multicolumn{1}{l}{PIGD} \\
    \midrule
    \multicolumn{1}{l}{\textbf{RF}} & \multicolumn{1}{l}{Thought disorder} & \multicolumn{1}{l}{Dressing} & \multicolumn{1}{l}{Hand pronate/supinate} & \multicolumn{1}{l}{Sleep disturbance} & Dressing & \multicolumn{1}{l}{Subtotal 2} & \multicolumn{1}{l}{Bradykinesia} \\
         &      & \multicolumn{1}{l}{Falling} &      & \multicolumn{1}{l}{Symptomatic orthostasis} & Hand pronate/supinate & \multicolumn{1}{l}{Subtotal 3} & \multicolumn{1}{l}{PIGD} \\
    \midrule
    \multicolumn{1}{l}{\textbf{LOGIT}} & \multicolumn{1}{l}{Thought disorder} & \multicolumn{1}{l}{Falling} & \multicolumn{1}{l}{Hand pronate/supinate} & \multicolumn{1}{l}{Sleep disturbance} & Dressing & \multicolumn{1}{l}{Subtotal 2} & \multicolumn{1}{l}{Bradykinesia} \\
         &      & \multicolumn{1}{l}{Freezing} & \multicolumn{1}{l}{Leg agility} & \multicolumn{1}{l}{Symptomatic orthostasis} & Speech & \multicolumn{1}{l}{Subtotal 3} &  \\
         &      & \multicolumn{1}{l}{Dressing} &      & \multicolumn{1}{l}{Early morning dystonia} & Hand pronate/supinate &      &  \\
         &      & \multicolumn{1}{l}{Handwriting} &      &      & Sleep disturbance &      &  \\
         &      &      &      &      & Symptomatic orthostasis &      &  \\
    \midrule
         &      &      &      &      &      &      &  \\
    \multicolumn{8}{c}{(b) At 12-month period} \\
    \midrule
    \rowcolor[rgb]{ .949,  .949,  .949} \multicolumn{1}{l}{\textbf{Method}} & \multicolumn{1}{l}{\textbf{UPDRS I }} & \multicolumn{1}{l}{\textbf{UPDRS II}} & \multicolumn{1}{l}{\textbf{UPDRS III}} & \multicolumn{1}{l}{\textbf{UPDRS IV}} & \textbf{All UPDRS items} & \multicolumn{1}{l}{\textbf{UPDRS subtotals}} & \multicolumn{1}{l}{\textbf{Composite measures}} \\
    \midrule
    \midrule
    \multicolumn{1}{l}{\textbf{DT}} & \multicolumn{1}{l}{Thought disorder} & \multicolumn{1}{l}{Falling} & \multicolumn{1}{l}{Action tremor} & \multicolumn{1}{l}{Sleep disturbance} & Dressing & \multicolumn{1}{l}{Subtotal 1} & \multicolumn{1}{l}{Tremor} \\
         &      & \multicolumn{1}{l}{Dressing} & \multicolumn{1}{l}{Hand pronate/supinate} & \multicolumn{1}{l}{Symptomatic orthostasis} & Hand pronate/supinate & \multicolumn{1}{l}{Subtotal 2} & \multicolumn{1}{l}{Bradykinesia} \\
    \midrule
    \multicolumn{1}{l}{\textbf{RF}} & \multicolumn{1}{l}{Thought disorder} & \multicolumn{1}{l}{Falling} & \multicolumn{1}{l}{Tremor at rest} & \multicolumn{1}{l}{Sleep disturbance} & Dressing & \multicolumn{1}{l}{Subtotal 2} & \multicolumn{1}{l}{Tremor} \\
         & \multicolumn{1}{l}{Motivation/Initiative} & \multicolumn{1}{l}{Dressing} & \multicolumn{1}{l}{Hand pronate/supinate} & \multicolumn{1}{l}{Symptomatic orthostasis} & Falling & \multicolumn{1}{l}{Subtotal 3} & \multicolumn{1}{l}{Bradykinesia} \\
         &      &      &      &      & Hand pronate/supinate &      &  \\
    \midrule
    \multicolumn{1}{l}{\textbf{LOGIT}} & \multicolumn{1}{l}{Thought disorder} & \multicolumn{1}{l}{Falling} & \multicolumn{1}{l}{Hand pronate/supinate} & \multicolumn{1}{l}{Dyskinesia (disability)} & Dressing & \multicolumn{1}{l}{Subtotal 2} & \multicolumn{1}{l}{Bradykinesia} \\
         &      & \multicolumn{1}{l}{Freezing} & \multicolumn{1}{l}{Leg agility} & \multicolumn{1}{l}{Anorexia, nausea, vomiting} & Speech &      &  \\
         &      & \multicolumn{1}{l}{Dressing} &      & \multicolumn{1}{l}{Sleep disturbance} & Hand pronate/supinate &      &  \\
         &      & \multicolumn{1}{l}{Walking} &      & \multicolumn{1}{l}{Symptomatic orthostasis} & Sleep disturbance &      &  \\
         &      &      &      &      & Symptomatic orthostasis &      &  \\
    \bottomrule
    \end{tabular}%
}
  \label{tab:addlabel}%
\end{table}%
\end{landscape}

\section {Discussion}
\label{Sec:discussion}
%\justify
This study demonstrated different ways of utilizing the UPDRS measurements for classifying people at the early stages of PD into fallers/non-fallers: using the individual items of the UPDRS, and aggregate measures derived from them. It is shown that the selected individual items of the UPDRS were more informative than the aggregate measures.  

It is worth noting that although UPDRS IV was often overlooked in similar studies for falls prediction, the odds ratio for falls prediction using symptomatic orthostasis and sleep disturbance  were greater than 1 in the univariate model. This is also confirmed by a comparably high purity of the classification results using UPDRS IV model as shown in Figure \ref{fig:DT_6m_items}(c). Yet, it has a relatively higher reduction in the classification rates compared to UPDRS II and UPDRS III when applied to cross-validation data.   

There is no clear difference between the performance of the methods used in this paper, with regards to the classification rates. Logistic regression provided higher accuracy in some models, yet there were cases where tree-based methods provide higher sensitivity, and vice versa. Logistic regression is attractive for its odds ratio interpretation for the effect, or contribution, of predictor variables in prediction. It is also less prone to over-fitting given the variables in the model are appropriately selected. Nevertheless, decision trees are more appealing since their visualization is easily interpretable, without the need for further calculation. The non-parametric approach of decision trees also offer the flexibility to handle a large number of variables, as demonstrated in this paper for models with UDPSR items as the predictor variables. While DT is often regarded to be more liable to over-fit, comparing the classification rates and selected variables with RF, the more robust method, the results are not greatly different for our data. Thus, we prefer and presented the fall/non-fall prediction rules based on DT, as in Section \ref{sec:riskfactors}.

Models were fit at two follow-up times to assess the effect of time to fall/non-fall prediction. The classification rates were varied at the two times, but the differences were not significant. There were also variations in variables being selected for the two time periods, but the differences were minor. So, overall it seemed that there was little change over time. This may be due to the relatively short differences in follow up times. However, on another perspective, this might imply that a shorter study time (6 months) could provide similar information than a longer study time (12 months). 

Regarding the measurements, the Movement Disorder Society (MDS) has released the improved version of the UPDRS called the MDS sponsored UPDRS (MDS-UPDRS). Yet, the  data used in this study were based on the UPDRS measurements. However, a reasonably high classification rates obtained showed that the UPDRS provided a useful information for fall/non-fall prediction. Moreover, results from models using only the UPDRS were comparable to those using additional information from other instruments (results not shown); suggesting that numerous measurements from many different instruments are not needed when information from a few particular instruments is used in an optimal way.

\section{Summary}
\label{Sec:summary}
%\justify
Through this study, we have provided empirical evidence that in the early stages of PD, fall/non-fall occurrences were better explained using  items of the UPDRS than using the composite measures. The highest classification rates for this model are: $80\%$ accuracy, $85\%$ sensitivity, and $77\%$ specificity, higher than previous studies. 

Among the four parts of the UPDRS, selected items from UPDRS Parts II and III produce a reasonably high classification rates compared to the other parts. The classification rates from all 4 methods at 2 time periods for UPDRS II items varied within these range: $70 - 75\%$  accuracy, $70-75\%$ sensitivity, and $68-71\%$ specificity. While for UPDRS III items, the range for the classification rates are $71-79\%$ accuracy, $73-83\%$ sensitivity, and $58-74\%$ specificity.  

We also identified variables that best predict fall/non-fall. It was also inferred that results from a 6-month follow-up time were not greatly different to that from a 12-month follow-up time, suggesting a shorter study time (6 months) could replace the longer study time (12 months).

Identification of the UPDRS items that are highly associated with falls offers several advantages. From a practical point of view, adjustments to treatment might be developed for PD patients to prevent falls. 
Focusing assessment based on the identified risk factors may provide more reliable responses, which will be advantageous for building a more informative model.
 
%\section{Bibliography} 
\bibliography{Manuscript_Assesing_the_predictability_of_UPDRS_Oct19}

\pagebreak
\appendix
\section{Falls odds ratio for the univariate logistic regression model at the 12-month of follow up. }
\label{appendixOR}

\begin{figure}[hp]
\centering\includegraphics[width=0.6\linewidth]{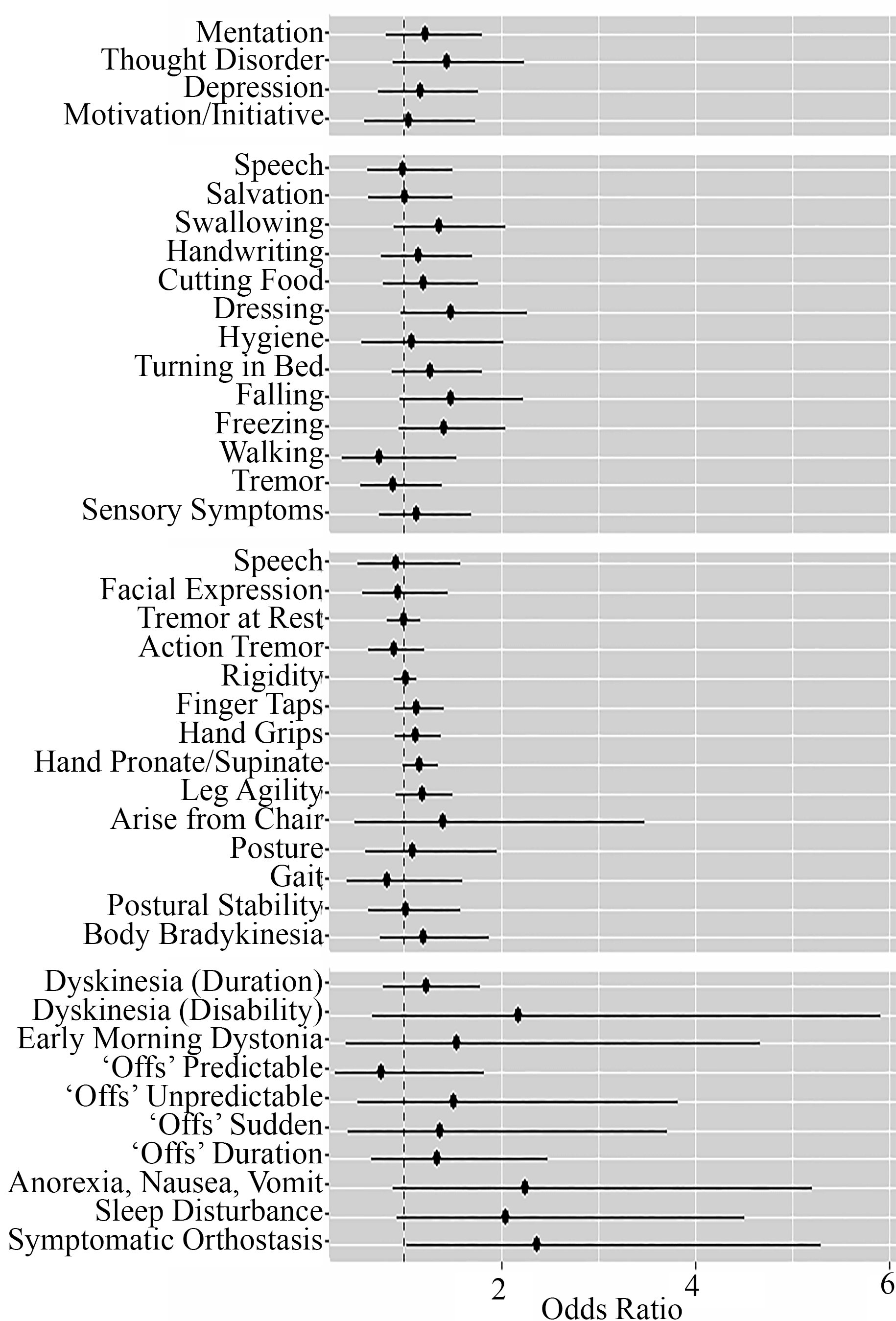}
\caption{Odds ratio (with 95\% CI) for the univariate logistic regression model using the UPDRS individual items as the explanatory variable. }
\label{fig:pic1}
\end{figure}

\begin{figure}[h]
\centering\includegraphics[width=0.6\linewidth]{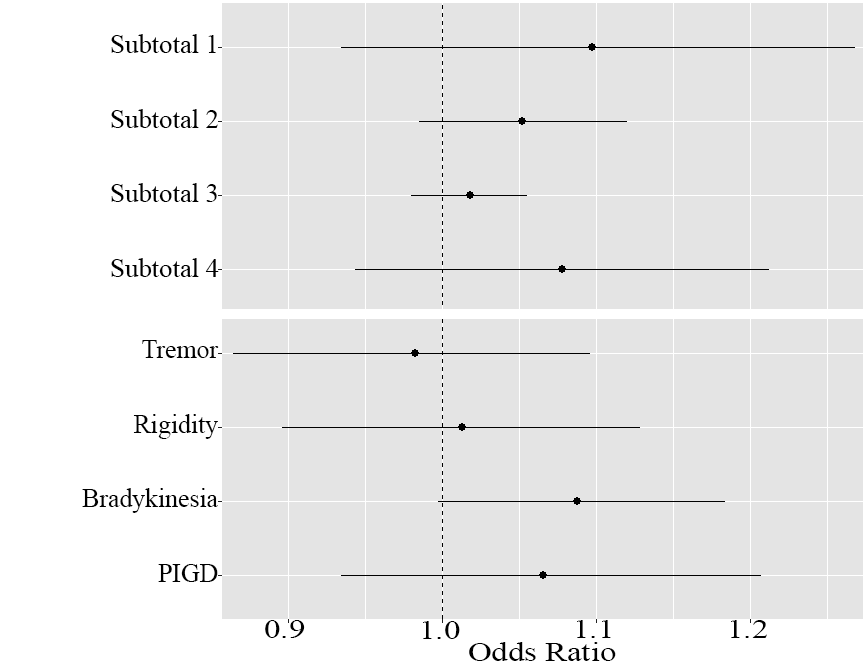}
\caption{Odds ratio (with 95\% CI) of falls classification at 12-month of follow-up, using univariate logistic regression model with aggregate measures of the UPDRS as the explanatory variable. Subtotals 1-4 are the sums of item scores in UPDRS Parts I - IV, respectively.}
\label{fig:ORaggregate_12m}
\end{figure}

\pagebreak
\section{ROC at 12 month of follow-up}

\begin{figure*}[htbp]
%\label{figA4}
\centering
\begin{subfigure}[b]{.5\textwidth}
  \centering
  \includegraphics[keepaspectratio=true,scale=0.4]{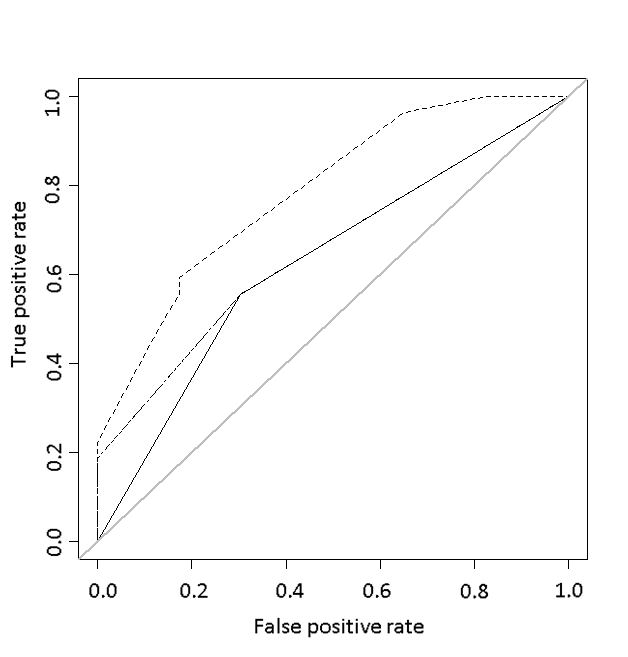}
  \caption{UPDRS I model.}
  \label{roc1}
\end{subfigure}%
~
\begin{subfigure}[b]{.5\textwidth}
  \centering
  \includegraphics[keepaspectratio=true,scale=0.4]{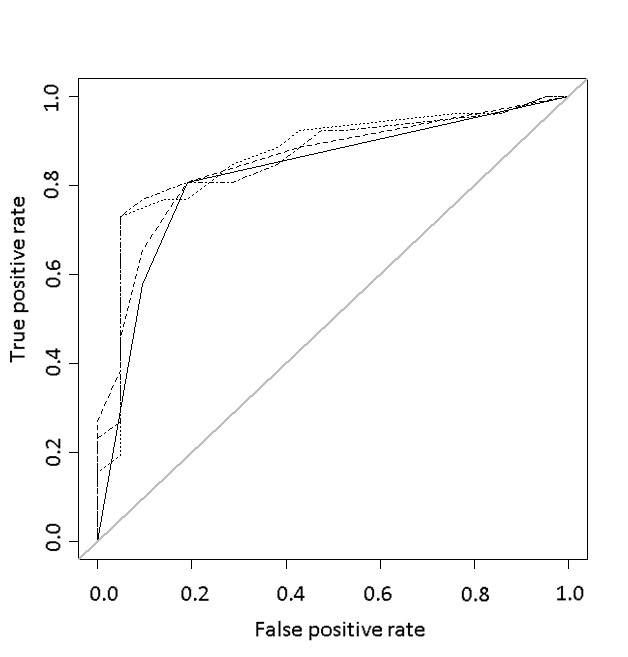}
  \caption{UPDRS II model.}
  \label{roc2}
\end{subfigure}%
~

\begin{subfigure}[b]{.5\textwidth}
  \centering
  \includegraphics[keepaspectratio=true,scale=0.4]{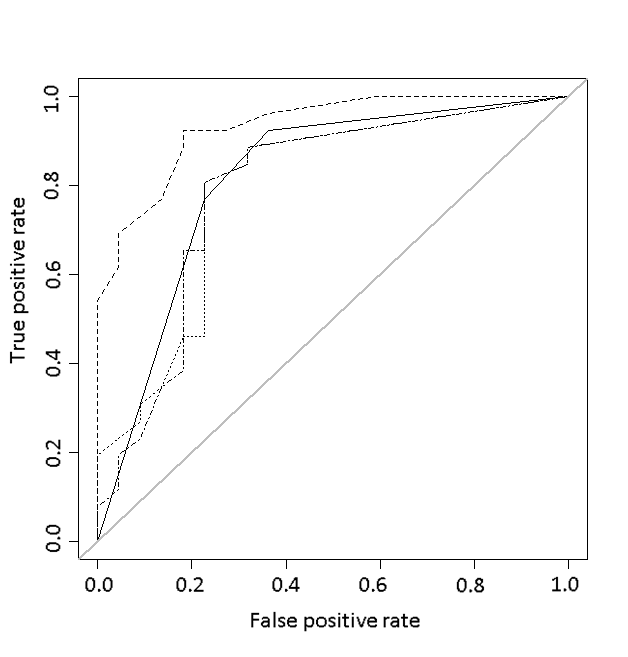}
  \caption{UPDRS III model.}
  \label{roc3}
\end{subfigure}%
~
\begin{subfigure}[b]{.5\textwidth}
  \centering
  \includegraphics[keepaspectratio=true,scale=0.4]{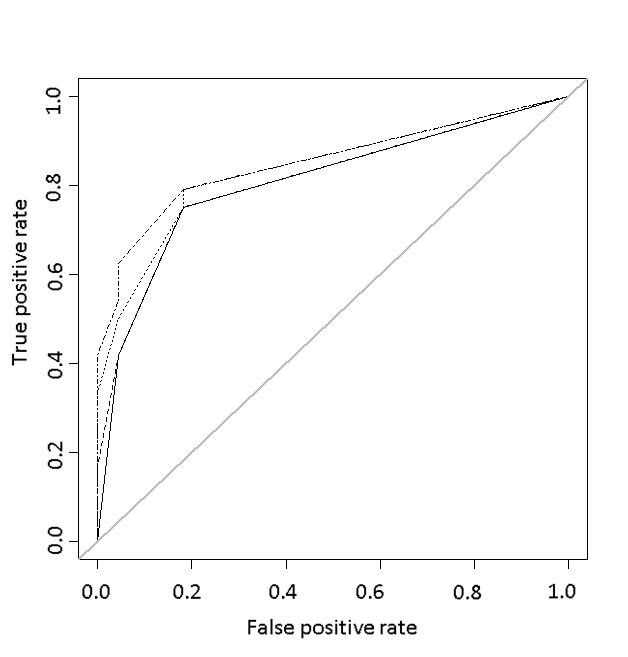}
  \caption{UPDRS IV model.}
  \label{roc4}
\end{subfigure}
\caption{ROC curves for falls classification at 12 month of follow-up using items of each part of the UPDRS. Classification methods employed are: Decision Tree (solid), Random Forest (dashed), logistic regression with stepwise (dotted), and logistic regression with BMA (dashed dotted).}
\label{fig:rocpart12m}
\end{figure*}

\begin{figure*}[htbp]
%\label{figA4}
\centering
\begin{subfigure}[b]{.3\textwidth}
  \centering
  \includegraphics[keepaspectratio=true,scale=0.28]{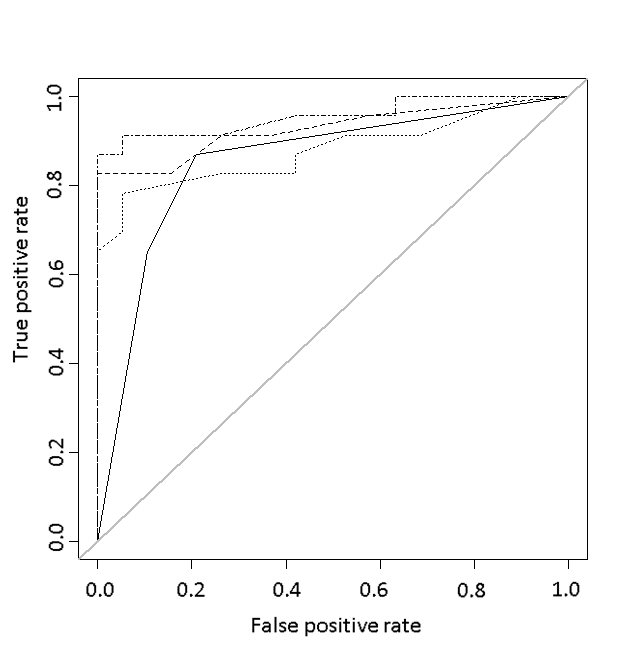}
  \caption{UPDRS model.}
  \label{roc5}
\end{subfigure}%
~
\begin{subfigure}[b]{.3\textwidth}
  \centering
  \includegraphics[keepaspectratio=true,scale=0.28]{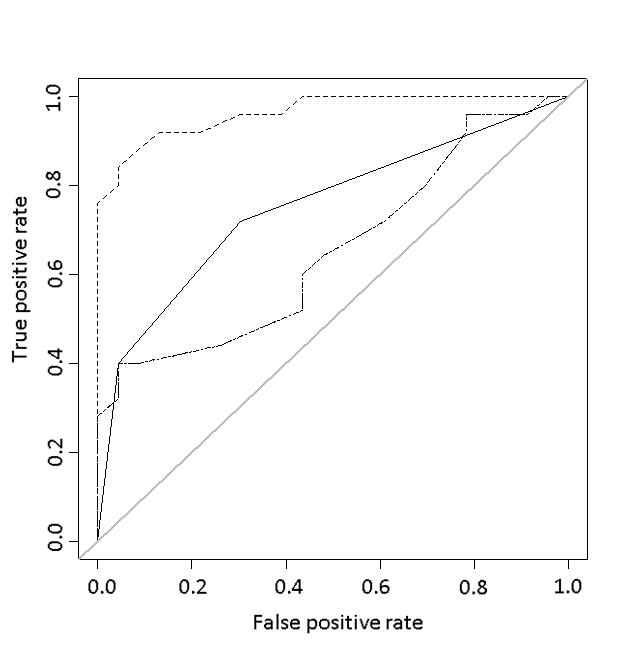}
  \caption{\textit{Subtotal} model.}
  \label{roc6}
\end{subfigure}%
~
\begin{subfigure}[b]{.3\textwidth}
  \centering
  \includegraphics[keepaspectratio=true,scale=0.28]{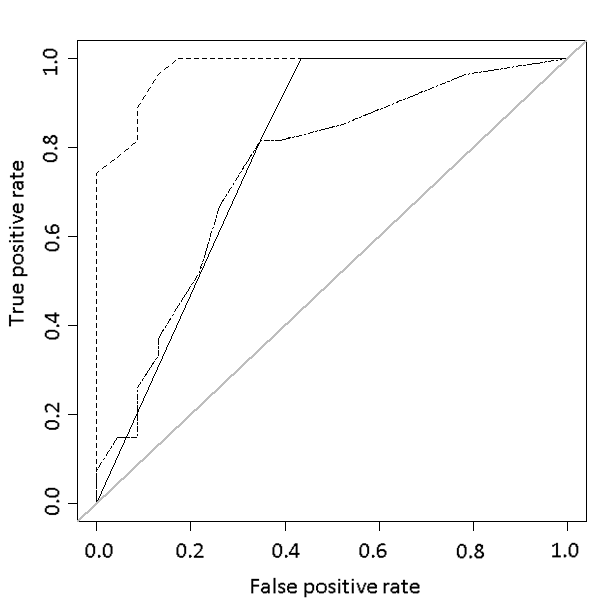}
  \caption{\textit{Composite} model.}
  \label{roc7}
\end{subfigure}%
\caption{ROC curves or falls classification at 12 month of follow-up using individual items (a), composite measures (b) and individual items and composite measures of the UPDRS. Classification methods employed are: Decision Tree (solid), Random Forest (dashed), logistic regression with stepwise (dotted), and logistic regression with BMA (dashed dotted).}
\label{fig:rocpart2_12m}
\end{figure*}
%%%%%%%%%%%%

\pagebreak
\section{Results of logistic regression with BMA}\label{sec:BMAresult}
No table is produced for UPDRS I model, as only item number 2, Thought disorder, was selected in the model for both 6-month and 12-month period. Also for Composite model, no table is produced as the only variable chosen for predictor is Bradykinesia, both at 6-month and 12-month periods.

\begin{table}[htbp]
  \centering
  \caption{BMA results for logistic regression using UPDRS Part II items as predictor variables (UPDRS II models).}
\scalebox{0.9}{% Table generated by Excel2LaTeX from sheet 'Sheet2'
    \begin{tabular}{lcccccrlccccc}
    \multicolumn{6}{c}{(a) At 6-month period} &      & \multicolumn{6}{c}{(b) At 12-month period} \\
\cmidrule{1-6}\cmidrule{8-13}    \rowcolor[rgb]{ .949,  .949,  .949}      & \multicolumn{5}{c}{Model}        & \cellcolor[rgb]{ 1,  1,  1}  &      & \multicolumn{5}{c}{Model} \\
\cmidrule{2-6}\cmidrule{9-13}    \rowcolor[rgb]{ .949,  .949,  .949}      & 1    & 2    & 3    & 4    & 5    & \cellcolor[rgb]{ 1,  1,  1}  &      & 1    & 2    & 3    & 4    & 5 \\
\cmidrule{1-6}\cmidrule{8-13}    Falling  & \cellcolor[rgb]{ .651,  .651,  .651}  & \cellcolor[rgb]{ .651,  .651,  .651}  &      &      & \cellcolor[rgb]{ .651,  .651,  .651}  &      & Falling  & \cellcolor[rgb]{ .502,  .502,  .502}  &      &      & \cellcolor[rgb]{ .502,  .502,  .502}  & \cellcolor[rgb]{ .502,  .502,  .502}  \\
    Freezing & \cellcolor[rgb]{ .651,  .651,  .651}  &      & \cellcolor[rgb]{ .651,  .651,  .651}  & \cellcolor[rgb]{ .651,  .651,  .651}  &      &      & Freezing & \cellcolor[rgb]{ .502,  .502,  .502}  & \cellcolor[rgb]{ .502,  .502,  .502}  & \cellcolor[rgb]{ .502,  .502,  .502}  & \cellcolor[rgb]{ .502,  .502,  .502}  &  \\
    Dressing &      & \cellcolor[rgb]{ .651,  .651,  .651}  &      &      &      &      & Dressing &      & \cellcolor[rgb]{ .502,  .502,  .502}  &      &      & \cellcolor[rgb]{ .502,  .502,  .502}  \\
    Handwriting &      &      & \cellcolor[rgb]{ .651,  .651,  .651}  &      &      &      & Walking & \cellcolor[rgb]{ .502,  .502,  .502}  & \cellcolor[rgb]{ .502,  .502,  .502}  & \cellcolor[rgb]{ .502,  .502,  .502}  &      & \cellcolor[rgb]{ .502,  .502,  .502}  \\
\cmidrule{1-6}\cmidrule{8-13}    \rowcolor[rgb]{ .949,  .949,  .949} Weight & 0.27 & 0.23 & 0.20 & 0.14 & 0.13 & \cellcolor[rgb]{ 1,  1,  1}  & Weight & 0.19 & 0.16 & 0.13 & 0.11 & 0.11 \\
\cmidrule{1-6}\cmidrule{8-13}    
\end{tabular}%
}
	\label{tab:bma.m2}%
\end{table}%
\begin{table}[htbp]
  \centering
  \caption{BMA results for logistic regression using UPDRS Part III items as predictor variables (UPDRS III models).}
\scalebox{0.7}{
    \begin{tabular}{lcccrlcc}
    \multicolumn{4}{c}{(a) At 6-month period} &      & \multicolumn{3}{c}{(b) At 12-month period} \\
\cmidrule{1-4}\cmidrule{6-8}    \rowcolor[rgb]{ .949,  .949,  .949}      & \multicolumn{3}{c}{Model} & \cellcolor[rgb]{ 1,  1,  1}  &      & \multicolumn{2}{c}{Model} \\
\cmidrule{2-4}\cmidrule{7-8}    \rowcolor[rgb]{ .949,  .949,  .949}      & 1    & 2    & 3    & \cellcolor[rgb]{ 1,  1,  1}  &      & 1    & 2 \\
\cmidrule{1-4}\cmidrule{6-8}    Hand pronate/supinate & \cellcolor[rgb]{ .651,  .651,  .651}  & \cellcolor[rgb]{ .651,  .651,  .651}  &      &      & Hand pronate/supinate & \cellcolor[rgb]{ .651,  .651,  .651}  &  \\
    Leg agility & \cellcolor[rgb]{ .651,  .651,  .651}  &      & \cellcolor[rgb]{ .651,  .651,  .651}  &      & Leg agility &      & \cellcolor[rgb]{ .651,  .651,  .651}  \\
\cmidrule{1-4}\cmidrule{6-8}    \rowcolor[rgb]{ .949,  .949,  .949} Weight & 0.50 & 0.47 & 0.04 & \cellcolor[rgb]{ 1,  1,  1}  & Weight & 0.81 & 0.10 \\
\cmidrule{1-4}\cmidrule{6-8}    \end{tabular}%
}
	\label{tab:bma.m3}%
\end{table}%
\begin{table}[htbp]
  \centering
  \caption{BMA results for logistic regression using UPDRS Part IV items as predictor variables (UPDRS IV models).}
\scalebox{0.7}{
    \begin{tabular}{lcccrlccccc}
    \multicolumn{4}{c}{(a) At 6-month period} &      & \multicolumn{6}{c}{(b) At 12-month period} \\
\cmidrule{1-4}\cmidrule{6-11}    \rowcolor[rgb]{ .949,  .949,  .949}      & \multicolumn{3}{c}{Model} & \cellcolor[rgb]{ 1,  1,  1}  &      & \multicolumn{5}{c}{Model} \\
\cmidrule{2-4}\cmidrule{7-11}    \rowcolor[rgb]{ .949,  .949,  .949}      & 1    & 2    & 3    & \cellcolor[rgb]{ 1,  1,  1}  &      & 1    & 2    & 3    & 4    & 5 \\
\cmidrule{1-4}\cmidrule{6-11}    Sleep disturbance & \cellcolor[rgb]{ .651,  .651,  .651}  &      & \cellcolor[rgb]{ .651,  .651,  .651}  &      & Sleep disturbance & \cellcolor[rgb]{ .502,  .502,  .502}  & \cellcolor[rgb]{ .502,  .502,  .502}  &      &      & \cellcolor[rgb]{ .502,  .502,  .502}  \\
    Symptomatic orthostasis & \cellcolor[rgb]{ .651,  .651,  .651}  & \cellcolor[rgb]{ .651,  .651,  .651}  &      &      & Symptomatic orthostasis & \cellcolor[rgb]{ .502,  .502,  .502}  & \cellcolor[rgb]{ .502,  .502,  .502}  & \cellcolor[rgb]{ .502,  .502,  .502}  & \cellcolor[rgb]{ .502,  .502,  .502}  &  \\
    Early morning dystonia &      & \cellcolor[rgb]{ .651,  .651,  .651}  &      &      & Dyskinesia (disability) &      &      & \cellcolor[rgb]{ .502,  .502,  .502} \textcolor[rgb]{ 1,  0,  0}{} &      &  \\
         &      &      &      &      & Anorexia, nausea, vomiting & \cellcolor[rgb]{ .502,  .502,  .502}  &      & \cellcolor[rgb]{ .502,  .502,  .502} \textcolor[rgb]{ 1,  0,  0}{} & \cellcolor[rgb]{ .502,  .502,  .502}  & \cellcolor[rgb]{ .502,  .502,  .502}  \\
\cmidrule{1-4}\cmidrule{6-11}    \rowcolor[rgb]{ .949,  .949,  .949} Weight & 0.76 & 0.17 & 0.08 & \cellcolor[rgb]{ 1,  1,  1}  & Weight & 0.35 & 0.18 & 0.17 & 0.10 & 0.06 \\
\cmidrule{1-4}\cmidrule{6-11}    \end{tabular}%
}
	\label{tab:addlabel}%
\end{table}%
\begin{table}[htbp]
  \centering
  \caption{BMA results for logistic regression using all items of the UPDRS as predictor variables (UPDRS models).}
\scalebox{0.7}{
    \begin{tabular}{lcccccrlccccc}
    \multicolumn{6}{c}{(a) At 6-month period} &      & \multicolumn{6}{c}{(b) At 12-month period} \\
\cmidrule{1-6}\cmidrule{8-13}    \rowcolor[rgb]{ .949,  .949,  .949}      & \multicolumn{5}{c}{Model}        & \cellcolor[rgb]{ 1,  1,  1}  &      & \multicolumn{5}{c}{Model} \\
\cmidrule{2-6}\cmidrule{9-13}    \rowcolor[rgb]{ .949,  .949,  .949}      & 1    & 2    & 3    & 4    & 5    & \cellcolor[rgb]{ 1,  1,  1}  &      & 1    & 2    & 3    & 4    & 5 \\
\cmidrule{1-6}\cmidrule{8-13}    Dressing & \cellcolor[rgb]{ .651,  .651,  .651}  & \cellcolor[rgb]{ .651,  .651,  .651}  &      &      & \cellcolor[rgb]{ .651,  .651,  .651}  &      & Dressing & \cellcolor[rgb]{ .651,  .651,  .651}  & \cellcolor[rgb]{ .651,  .651,  .651}  &      &      & \cellcolor[rgb]{ .651,  .651,  .651}  \\
    Speech &      &      &      &      & \cellcolor[rgb]{ .651,  .651,  .651}  &      & Speech &      &      &      &      & \cellcolor[rgb]{ .651,  .651,  .651}  \\
    Hand pronate/supinate & \cellcolor[rgb]{ .651,  .651,  .651}  &      &      & \cellcolor[rgb]{ .651,  .651,  .651}  & \cellcolor[rgb]{ .651,  .651,  .651}  &      & Hand pronate/supinate & \cellcolor[rgb]{ .651,  .651,  .651}  &      &      & \cellcolor[rgb]{ .651,  .651,  .651}  & \cellcolor[rgb]{ .651,  .651,  .651}  \\
    Sleep disturbance &      &      & \cellcolor[rgb]{ .651,  .651,  .651}  &      &      &      & Sleep disturbance &      &      & \cellcolor[rgb]{ .651,  .651,  .651}  &      &  \\
    Symptomatic orthostasis & \cellcolor[rgb]{ .651,  .651,  .651}  & \cellcolor[rgb]{ .651,  .651,  .651}  & \cellcolor[rgb]{ .651,  .651,  .651}  & \cellcolor[rgb]{ .651,  .651,  .651}  &      &      & Symptomatic orthostasis & \cellcolor[rgb]{ .651,  .651,  .651}  & \cellcolor[rgb]{ .651,  .651,  .651}  & \cellcolor[rgb]{ .651,  .651,  .651}  & \cellcolor[rgb]{ .651,  .651,  .651}  &  \\
\cmidrule{1-6}\cmidrule{8-13}    \rowcolor[rgb]{ .949,  .949,  .949} Weight & 0.25 & 0.20 & 0.17 & 0.13 & 0.11 & \cellcolor[rgb]{ 1,  1,  1}  & Weight & 0.54 & 0.19 & 0.12 & 0.12 & 0.04 \\
\cmidrule{1-6}\cmidrule{8-13}    \end{tabular}%
}
  \label{tab:addlabel}%
\end{table}%
\begin{table}[htbp]
  \centering
  \caption{BMA results for logistic regression using subtotals of the UPDRS as predictor variables (\textit{Subtotal} models) at 6-month period. At 12-month, only Subtotal 2 was selected to include in the model.}
    \begin{tabular}{lccc}
   % \multicolumn{4}{c}{(a) At 6 month period} \\
    \midrule
    \rowcolor[rgb]{ .949,  .949,  .949}      & \multicolumn{3}{c}{Model} \\
\cmidrule{2-4}    \rowcolor[rgb]{ .949,  .949,  .949}      & \multicolumn{1}{r}{1} & \multicolumn{1}{r}{2} & \multicolumn{1}{r}{3} \\
    \midrule
    \midrule
    Subtotal 2 & \cellcolor[rgb]{ .651,  .651,  .651}  & \cellcolor[rgb]{ .651,  .651,  .651}  &  \\
    Subtotal 3 &      & \cellcolor[rgb]{ .651,  .651,  .651}  & \cellcolor[rgb]{ .651,  .651,  .651}  \\
    \midrule
    \rowcolor[rgb]{ .949,  .949,  .949} Weight & 0.89 & 0.06 & 0.05 \\
    \bottomrule
    \end{tabular}%
  \label{tab:addlabel}%
\end{table}%

\end{document}